\documentclass[a4paper,10pt]{article}
\usepackage[pdftex]{graphicx}
\PassOptionsToPackage{hyphens}{url}\usepackage{hyperref}
\hypersetup{
     colorlinks = true,
     allcolors = blue
}
\usepackage{amsmath,amsfonts,amssymb}
\usepackage{fullpage}
\usepackage{authblk}
\usepackage{setspace}
\usepackage{subcaption}
\usepackage{booktabs}
\usepackage{tabularx}

\usepackage[explicit]{titlesec}
\usepackage{sidecap}
\usepackage{pbox}

\usepackage{placeins}
\usepackage{float}

\usepackage{amsfonts}
\usepackage{amsmath}
\usepackage{multirow}
\usepackage{longtable}
\usepackage{makecell}
\usepackage{graphicx}
\usepackage{xcolor}
\usepackage{comment}
\newlength{\figureDefaultWidth}
\newlength{\figureOneCol}
\newlength{\figureSmaller}
\newcommand{\SI}{in the \textit{Supplementary Information}}

\date{}
\title{Macroscopic properties of buyer-seller networks in online marketplaces}

\author[a]{Alberto Bracci}
\author[b]{J\"{o}rn Boehnke}
\author[c]{Abeer ElBahrawy} 
\author[d]{Nicola Perra} 
\author[e]{Alexander Teytelboym}  
\author[a,f,g,+]{Andrea Baronchelli}

\affil[a]{Department of
Mathematics, City, University of London, London EC1V 0HB, UK}
\affil[b]{Graduate School of
Management, University of California Davis, 1 Shields Ave Davis, CA, USA}
\affil[c]{Chainalysis Inc, NY, USA,}
\affil[d]{Networks and Urban Systems Centre, University of Greenwich, London SE10 9LS, UK,}
\affil[e]{Department of
Economics, University of Oxford, Oxford OX1 3UQ, UK}
\affil[f]{UCL Centre for Blockchain Technologies, University
College London, London WC1E 6BT, UK}
\affil[g]{The Alan Turing Institute, London NW1 2DB, UK}
\affil[+]{\small Corresponding Author: abaronchelli@turing.ac.uk}

\begin{document}
\maketitle

\begin{abstract}
\textbf{Online marketplaces are the main engines of legal and illegal e-commerce, yet    their empirical properties are poorly understood due to the absence of large-scale data  .
% yet the buyer-seller networks behind them are poorly understood 
We analyze two comprehensive datasets containing 245M transactions (16B USD) that took place on online marketplaces between 2010 and 2021, covering  28 dark web marketplaces, i.e., unregulated markets whose main currency is Bitcoin, and 144 product markets of one popular regulated e-commerce platform.
We show that transactions in online marketplaces exhibit strikingly similar patterns despite significant differences in language, lifetimes, products, regulation, and technology.
Specifically, we find remarkable regularities in the distributions of transaction amounts, number of transactions, inter-event times and time between first and last transactions.
We show that buyer behavior is affected by the memory of past interactions and use this insight to propose a model of network formation reproducing our main empirical observations.
Our findings have implications for understanding market power on online marketplaces as well as inter-marketplace competition, and provide empirical foundation for theoretical economic models of online marketplaces.}
\end{abstract}

\section*{Introduction}
Much of online trade happens on regulated and unregulated online marketplaces. 
Regulated online marketplaces include Amazon, Craigslist, eBay, Walmart, Alibaba (China), Rakuten (Japan), Gumtree (UK), and Mercado Libre (South America).
Unregulated online marketplaces, such as Silk Road, AlphaBay, and Hydra, that specialise in the sale of illicit goods, have proliferated on (and disappeared from) the dark web since the introduction of Bitcoin~\cite{specific_4,abeer_collectivedynamics,dwm_2,bracci2021dark}. %nadini2021emergence}. % bracci2021dark_after, 
The amount of transactions in online marketplaces is vast and growing. 
For example, in 2020 Amazon reported a net revenue of 386B USD~\cite{market_numbers3}, while in 2019 the ecosystem of dark web marketplaces (DWMs) had reached a total volume of 4B USD~\cite{abeer_collectivedynamics}.

Online marketplaces are commercial websites that allow participating buyers and sellers to exchange information about prices and products and to execute transactions ~\cite{market_definition_1,market_definition_2,market_definition_3}.
Sellers can usually post an ad for their product that includes a product description, a price and a shipping method. 
Buyers instead can see all relevant product ads matching search keywords, and have access to reviews and seller ratings. 
When a purchase is made, the payment is processed through the platform, while shipping is usually taken care of by the seller.

Despite the importance of online marketplaces for e-commerce and global trade~\cite{global_trade_1,global_trade_2},    little is known about their empirical properties, transaction patterns and the resulting buyer-seller networks.  
% little is known about aggregate properties of their buyer-seller networks. 
The properties of the transaction network could, however, provide important insights into the presence of market power~\cite{market_power_better_1,market_power_better_2}, the nature of inter-platform competition~\cite{competition_1,competition_2}, product design~\cite{product_design_1}, the effects of reputation on sellers' revenue growth~\cite{network_reputation_better_1}, and the long-run sustainability of the platforms~\cite{sustainability_better_1}. 
   Moreover, measuring properties of the buyer-seller networks could help provide empirical foundations for theoretical models of online marketplaces, from the estimation of model parameters to suggesting specific model mechanisms.  
However, buyer-seller networks in online marketplaces have specific features that make them different from other networks (e.g., social networks): they exhibit a naturally bipartite structure; most transactions (links) occur between anonymous agents; transaction activity might be infrequent and sporadic.
Moreover, the structure of buyer-seller networks could depend on the nature of the traded products, on the types of buyers and sellers, on the user experience on the marketplace, or even on the legal, institutional and geographic constraints.

One strand of prior work relevant to our paper has touched on various aspects of regulated online marketplaces.
For example, the role of reputation and feedback~\cite{market_numbers,reputation_2,negative_reputation} has been identified as one of the main drivers of the worldwide success of regulated online platforms~\cite{pref_attach_1}. 
Other work has looked at consumer search and the effect of rankings on product choice~\cite{consumer_search_1,consumer_search_2,consumer_search_3,consumer_search_4}, online auction markets~\cite{auction_1,auction_2,auction_3,auction_4} and price formation~\cite{price_1,price_2,price_3,price_4,price_5} in online markets (For a more complete but less up-to-date review see~\cite{market_definition_1}).
Another strand of research has studied unregulated marketplaces. This work has focused on country-specific studies~\cite{country_1,country_2,interview_2}, the effects of closures and law enforcement raids~\cite{closure_1,dwm_2,abeer_collectivedynamics,closure_2,nadini2022emergence}, the characterization of the trade of specific goods~\cite{specific_3,specific_5,specific_7}, the importance of  geography~\cite{specific_2,geography_1}, or sociological interview-based studies~\cite{interview_1,interview_2}. %,bracci2021dark_after}
However, most work on unregulated online marketplaces was limited to specific markets, and focused on information available from public listings (e.g., using crawled data)~\cite{closure_1,specific_2,specific_3,specific_4,specific_5,specific_7}. 

In this paper, we focus on patterns in transactions which typically cannot be publicly observed either on regulated or unregulated online marketplaces. 
We analyze two datasets. 
The first dataset contains 220M transactions between 99M buyers and 7.4M sellers which occurred in 144 randomly sampled product markets of one regulated e-commerce platform between 2010 and 2020, for a total volume of over 10B USD. 
The second dataset contains 25M transactions involving 17M entities with a total volume of 4.2B USD which occurred in 28 major DWMs between 2011 and 2021, for a total volume of 4.2B USD (for more details on the datasets see \textit{Materials and Methods}). 
In both cases, the datasets cover all transactions which occurred in each corresponding market.

We observe striking similarities in user behavior across online marketplaces, despite their significant differences. 
First, we find that the number of transactions, amount, inter-event time and time between first and last transaction are highly heterogeneous across users but follow consistent fat-tailed distributions across all marketplaces.
Then, we show that individual behavior is influenced by past purchases similarly (albeit less strongly) to what is observed in the renewal of past ties in social networks~\cite{karsai_time_2015,ubaldi_asymptotic_2016}.  
Finally, we propose a simple model of buyer-seller interactions that reproduces the main stylized facts of the data and emphasizes the critical role of preferential attachment~\cite{social_systems_1,social_systems_3} and memory in the market dynamics.

\section*{Results}
\label{sec:results}

\subsection*{Empirical properties of buyer-seller networks}
\label{sec:results1}

In order to characterize the buyer-seller networks, we start by analyzing different aggregate user-level quantities. 
First, we study the distributions for the number and amount of user transactions.
Results for DWMs and for each e-commerce market are shown in Fig.1a-d, where black and yellow lines are obtained by aggregating all users in the respective datasets. 
Single distributions display common behavior, spanning several orders of magnitudes. 
It is important to note that distributions are computed without any rescaling or manipulation of the data, and that higher values generally reached by the regulated platform in all distributions are exclusively due to the different platform sizes. 
The slight discrepancy between the distributions in the total number of received transactions can be ascribed to the different nature of the two datasets: while in the DWM dataset sellers can withdraw the earnings from several market trades at once, in the e-commerce data each transaction corresponds to a single purchase. 

We then analyze the temporal dimension of the data.
We focus on the distribution of user lifetimes, defined as the time between the first and last user transaction in the market, and the inter-event time between two successive transactions of the same user. 
Again, we find remarkable regularities across different DWMs and different regulated product markets, as shown in Fig.1e-h. 
In these distributions, as before, we also observe the effects of different sizes of marketplaces.
The similarity between different distributions is particularly pronounced in the meaningful timescales between an hour and a month/year. 
Discrepancies for longer periods are due to the different lifetimes of the markets, whereas discrepancies for shorter timescales can be explained by the different nature of the two datasets: precise timestamps on transaction data for the regulated marketplaces vs. times at which the transaction is actually added to the Bitcoin blockchain (which depends on its algorithm) for the DWM dataset.

\begin{figure*}[ht]
	\centering
	\includegraphics[width=16cm]{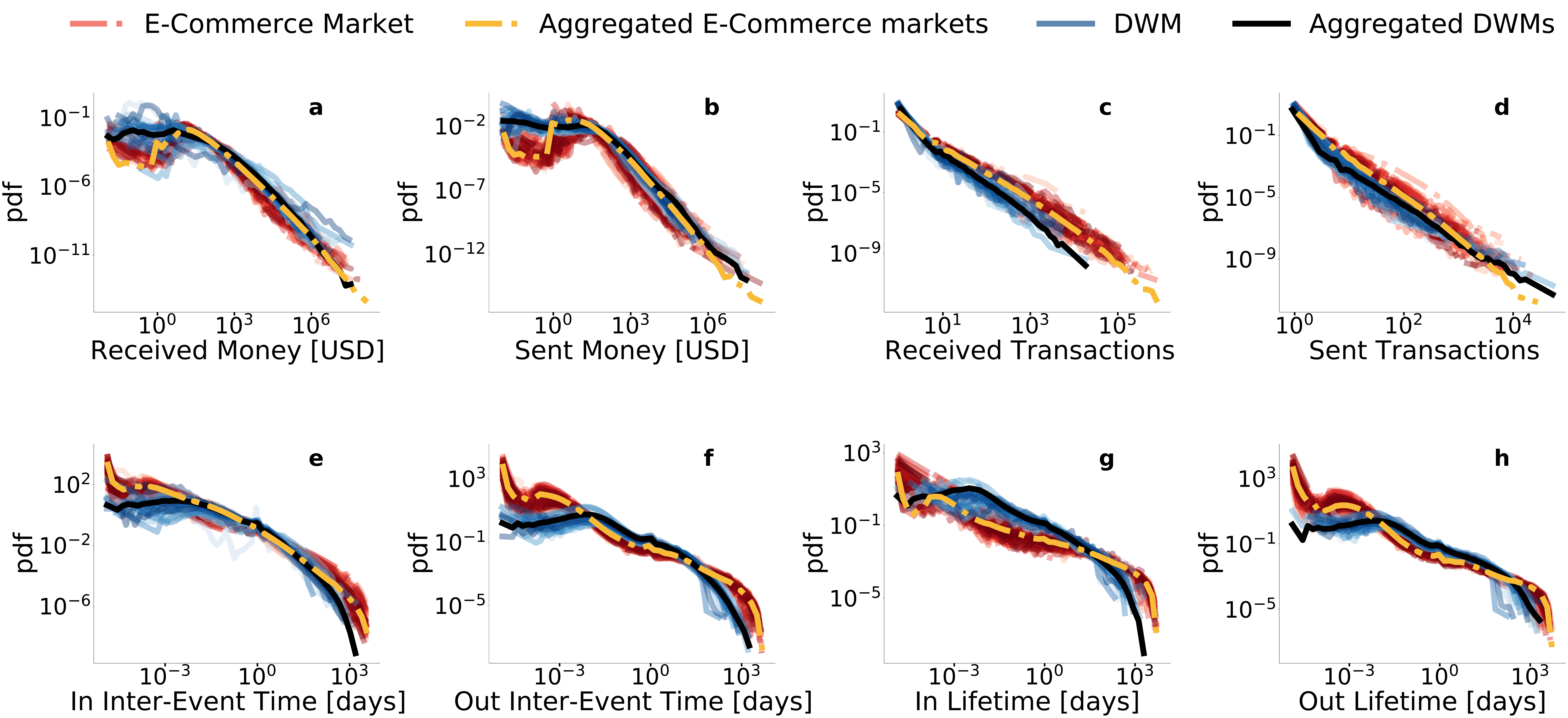}
	\caption{\label{fig:all_distributions}\textbf{Online marketplaces show strikingly similar patterns according to different aggregated quantities.} 
	Top (a to d): Distributions of 4 analyzed users aggregate quantities: money and number of transactions both sent and received. 
	Bottom (e to h): Distributions of 4 analyzed users temporal quantities: the inter-event time (time between successive transactions) and the lifetime (time between first and last transaction) both measured in days.
	Each blue line represents one DWM, the black line is the distribution built aggregating all DWMs together, each dashed red line represents one e-commerce platform market, while the dashed yellow line is the distribution aggregating all e-commerce markets.
	Similar patterns are observed between different markets in the same platform, but also across regulated and unregulated online marketplaces.}
\end{figure*}

Having considered buyers and sellers separately, we now investigate the dynamics of buyer-seller relationships and the evolution of the buyer-seller network. 
We limit this analysis to e-commerce markets, since DWMs data do not contain buyer-seller links (see \textit{Materials and Methods} for more details). 
We first consider how single users distribute their purchases across sellers: for example, buyers could purchase equally from multiple sellers or, alternatively, buyers could show loyalty to one seller from which they do most of their purchases. 
A standard way to quantify how distributed or concentrated this pattern is to compute the normalized entropy for the purchases of each buyer $i$ as in Eq.1{eq:entropy}, and then compute its distribution for all markets. 
The normalized entropy is defined as

\begin{equation}
\label{eq:entropy}
e_i =  - \sum_{j=1}^{J}n_i^j \log_2(n_i^j) / \log_2(J)
\end{equation}

\noindent where $n_i^j$ is the share of buyer $i$'s purchases from seller $j$ and we sum over the $J$ sellers the buyer purchased from. 
Fig.2a shows that the distributions are fat-tailed, with buyers populating the full $[0,1]$ support but with most of the mass towards the top, meaning that most buyers buy a similar number of times from the different sellers they purchase from.
Buyers with zero entropy, who buy from just one seller, were excluded from the figure for visual clarity, but these were almost exclusively buyers who only made a single purchase (see Fig.~S1 \SI ).

\begin{figure}[ht]
	\centering
	\includegraphics[width=16cm]{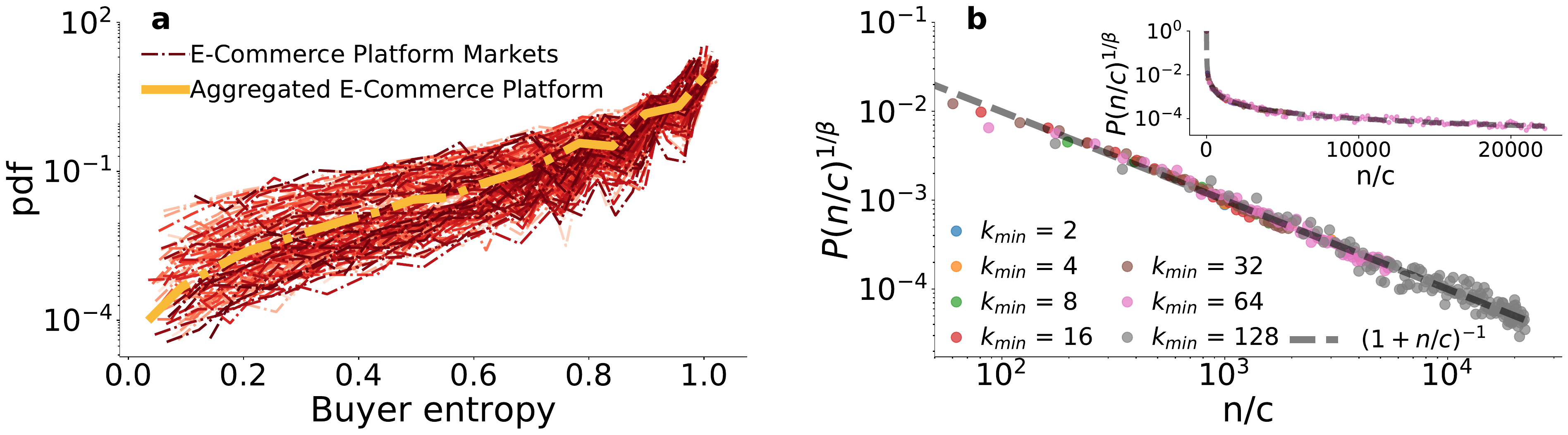}
	\caption{\label{fig:memory} \textbf{Buyer memory affects their purchase decisions.} 
	a) Normalised buyer entropy distribution for each e-commerce market (red), and the whole e-commerce platform (yellow), excluding users with zero entropy (mostly users with one transaction, see Fig.~S1 \SI) for visual clarity. 
	The fat-tailed distributions span the full $[0,1]$ range, with most buyers almost equally buying from multiple sellers. 
	b) $P(n)$ is the probability to buy from a new seller after a buyer has already bought from $n$ different ones. 
	Each degree class $k_{min} \leq k \leq 2 \cdot k_{min} - 1$ is rescaled according to the fitted value of $c$ and $\beta$ (see Tab.~S1 \SI for the values), with Eq.2 (dashed line) well reproducing the memory effect on the buyers' behavior: the more sellers they try, the less likely they are to buy from a new one.}
\end{figure}
 
The observed normalized entropy distributions are compatible with different kinds of temporal patterns produced by two possible choices: either buyers choose to engage with new sellers they have never purchased from (i.e., exploration) or they return to a known seller (i.e., exploitation). 
We investigate these dynamics by leveraging insight from the social networks literature, where several studies have investigated how users explore and exploit social connections by renewing previously activated ties or by establishing new ones~\cite{karsai_time_2015,ubaldi_asymptotic_2016}. 
Indeed, across different types of social networks, the temporal evolution of links that a person forms with their contacts can be captured by the following expression:

\begin{equation}
\label{eq:memory}
P(n) = \left(1+n/c(k_{min})\right)^{-\beta(k_{min})}
\end{equation}

\noindent where--- now using the language of online marketplaces ---$P(n)$ is the probability that a buyer (node) of degree $n$ (who has already bought from $n$ different sellers) chooses to buy from a new seller, while $c$ and $\beta$ are positive constants,depending on the final degree of the buyer, which measure their propensity to explore new sellers and thus the effect of memory. 
Following the procedure proposed in \cite{karsai_time_2015} (see \textit{Supplementary Information} for more details), we group nodes in different classes according to the final degree: a buyer is in class $k_{min}$ if the final degree $k$ satisfies $k_{min} \leq k \leq 2 \cdot k_{min} - 1 $, starting from $k_{min}=2$. 
We then fit~ Eq.ref{eq:memory} to each node class obtaining a value of $c(k_{min})$ and $\beta(k_{min})$ (see Tab.~S1 \SI). 

Results are shown in Fig.2b. Since different classes feature different values of $\beta$ and $c$, we plot a rescaled $P(x)^{1/\beta}$ as a function of $n/c$. 
Indeed, Eq.2 becomes $1/(1+x)$ (dashed line in Fig.~2b) for every degree class $k_{min}$, where $x=n/c$.
In other words, we re-scale both axes assuming the empirical behavior is captured by Eq.2. 
As shown in Tab.~S1 \SI, the parameter values are independent of the degree class and suggest a weaker ($\beta \sim 10^{-1}$) effect than previously observed in social networks ($0.48 \leq \beta \leq 2$)~\cite{ubaldi_asymptotic_2016}.
The close fit of the data to the predicted memory for different $k_{min}$ indicates the applicability of Eq.2  in the dynamics of buyer-seller relationships. 
While users have different propensities to explore new sellers, they follow the same mechanism: the more sellers a user has bought from, the less likely is their next purchase from a new seller. 

\subsection*{Modeling buyer-seller networks}
\label{sec:results3}

In order to understand possible mechanisms that drive the properties of buyer-seller networks, we propose an agent-based model aimed at capturing the patterns observed in the previous section. The main features of the model are:

\begin{enumerate}
\item \textbf{Activity.}  The rate at which buyers make transactions. 
As shown in Fig.1, in both e-commerce and DWMs buyers feature heterogeneous propensities to make purchases. 
\item \textbf{Memory.} When making new transactions buyers can either choose a seller they already bought from or pick a new one.
As shown in Fig.2b and discussed above, buyers have a memory of the sellers they had interacted with, and this memory affects their future purchases.
\item \textbf{Preferential attachment.} The attractiveness (i.e., popularity) of a seller is proportional to the number of their sales. 
This attractiveness captures the fact that, in online marketplaces, sellers are rated based on the feedback they receive from the buyers, and buyers prefer sellers with higher ratings, other things equal~\cite{reputation_2,market_numbers,reputation_dwm_eccdma,reputation_dwm_1,dwm_1}.
Here, we focus on the number of sales rather than sale volume to capture the fact that it is mainly frequency of transactions that matters for seller reputation. 
\end{enumerate}

Given these three ingredients the model dynamics is as follows. 
The system consists of $N$ buyers and $M$ sellers. 
At $t=0$ we assign the activity $a_i$ to each buyer $i$. Each seller $j$ starts with attractiveness $A_j = 1$. 
At each time step $t$, each buyer makes a purchase with probability $a_i \cdot \Delta_t$, where $\Delta_t$ is the simulation time step (fixed to 1 from now on).
A buyer who interacted with $n$ sellers in the past has probability $P(n) = \left(1+n/c)\right)^{-\beta}$ of choosing a new seller and $1-P(n)$ of returning to a known one.
In the first case, the buyer selects a new seller $j$ proportionally to their attractiveness~\cite{attractiveness} $A_j$, in the latter, the buyer selects it proportionally to the number of previous interactions. 
In other words, buyers select sellers either according to past purchases or to their popularity. 
In both cases the attractiveness of the seller is increased by $\mu$. 
This model produces a bipartite temporal network: at each time step $t$ we build a network in which two types of nodes--- buyers and sellers---are linked if the buyer has purchased from that seller at time $t$. 
These networks are then combined together in an aggregated network, where each buyer-seller link is weighed according to the number of purchases between that buyer and that seller across time.

Compared to other activity-driven models developed to capture the temporal evolution of different social networks \cite{karsai_time_2015, ubaldi_asymptotic_2016, perra_activity_2012}, our model extends the framework to bipartite networks of buyers and sellers and introduces the preferential attachment guiding the buyer selection process. 
Henceforth, we will refer to the model lacking preferential attachment, proposed in~\cite{karsai_time_2015}, as \textit{Model NoPA}. 
We will also consider a version of the model that does not include the memory element (\textit{Model NoMem}). 
Comparing these versions of the model will allow us to identify the role played by the different mechanisms.

A standard way to define and measure user activity in a (social) network is $a_i = n_i / \Sigma_\ell n_\ell$, where $n_i$ is the number of purchases made by buyer $i$, where the sum is over all buyers in their market. 
In Fig.3 we show the activity distributions of all e-commerce markets (a) and all DWMs (b). 
While curves exhibit fat-tailed behavior, they no longer overlap due to different activity ranges and shapes in different product markets.
As a result, we need to use market-specific empirical activity distributions as inputs for our model.

\begin{figure}[ht]
	\centering
	\includegraphics[width=16cm]{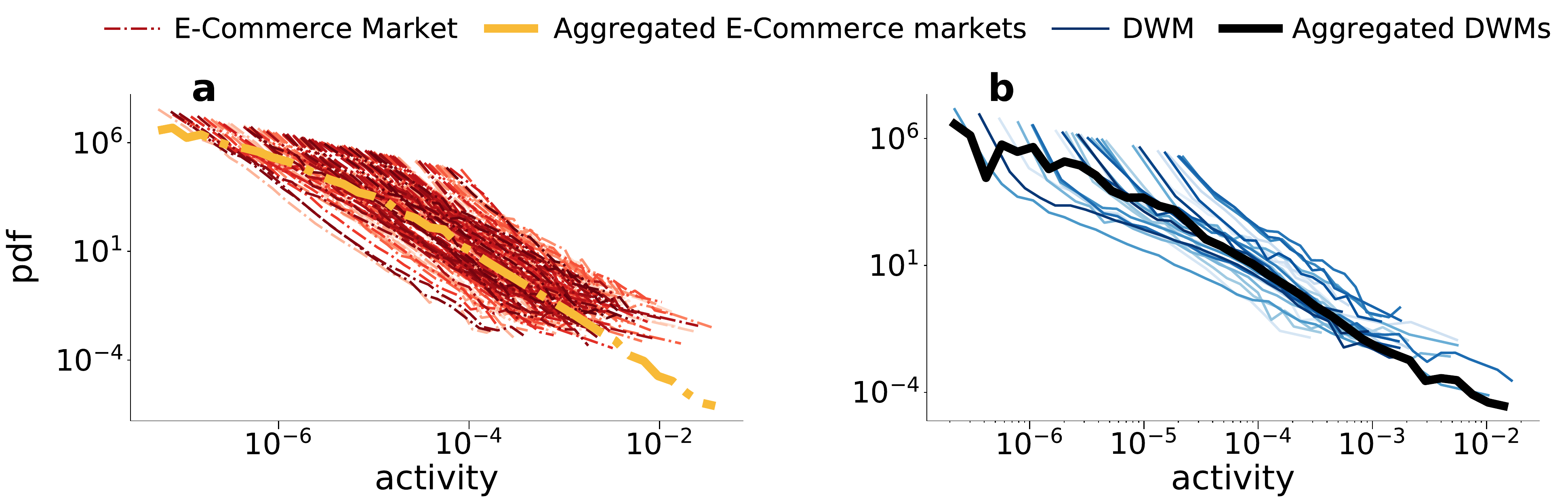}
	\caption{\label{fig:empirical_distributions} \textbf{Empirical activity distributions.} 
	a): Activity distribution for all e-commerce markets in red, activity distribution of all aggregated markets in yellow. 
	b): Activity distribution for all DWMs in blue, activity distribution of all aggregated DWMs in black.}
\end{figure}

We now fit the model to the e-commerce data. 
As mentioned above, since the DWM dataset does not contain the full bipartite buyer-seller network, we cannot test all the model predictions on the DWM data.
We employ a data-driven approach, fine-tuning the model to each single market so we can more faithfully compare the simulation results with the empirical buyer-seller networks. 
In the main text we show results for two different product markets, 26 more are shown in Fig.~S2-S5 \SI, for a total of 28 markets (see \textit{Supplementary Information} for the sampling procedure). 
We fix parameters $\beta=0.1$ and $c=0.001$ which we fitted previously (see Tab.~S1 \SI), and use the empirical activity distributions as measured in the data (see Fig.3a) to reflect the observed heterogeneity across different markets.
The value of $\mu$ is determined with Maximum Likelihood Estimation for each market (see \textit{Supplementary Information} for more details).

Results are in Fig.4. 
We first compare the model's output with the empirical distributions of the final seller attractiveness and degree. 
The attractiveness of a seller $j$ is their market share $A_j = s_j / \Sigma_\ell s_\ell$, where $s_j$ is the total number of sales of seller $j$  and the sum is over all the sellers. 
Fig.4 shows that the model  reproduces both distributions well, while the NoPA variation of our main model (without preferential attachment) fails to capture the heterogeneity (up to six orders of magnitude) of these curves, emphasizing how preferential attachment is key to reproducing the presence of very active sellers.
We then consider the buyer side of the network. 
We first study the degree distribution. 
Fig.4 shows that the model captures the empirical distributions, while the absence of buyer memory generally leads to a small overestimation of the tails, since it does not induce the repetition of past interactions with a subset of buyers. 

\begin{figure*}[ht]
	\centering
	\includegraphics[width=16cm]{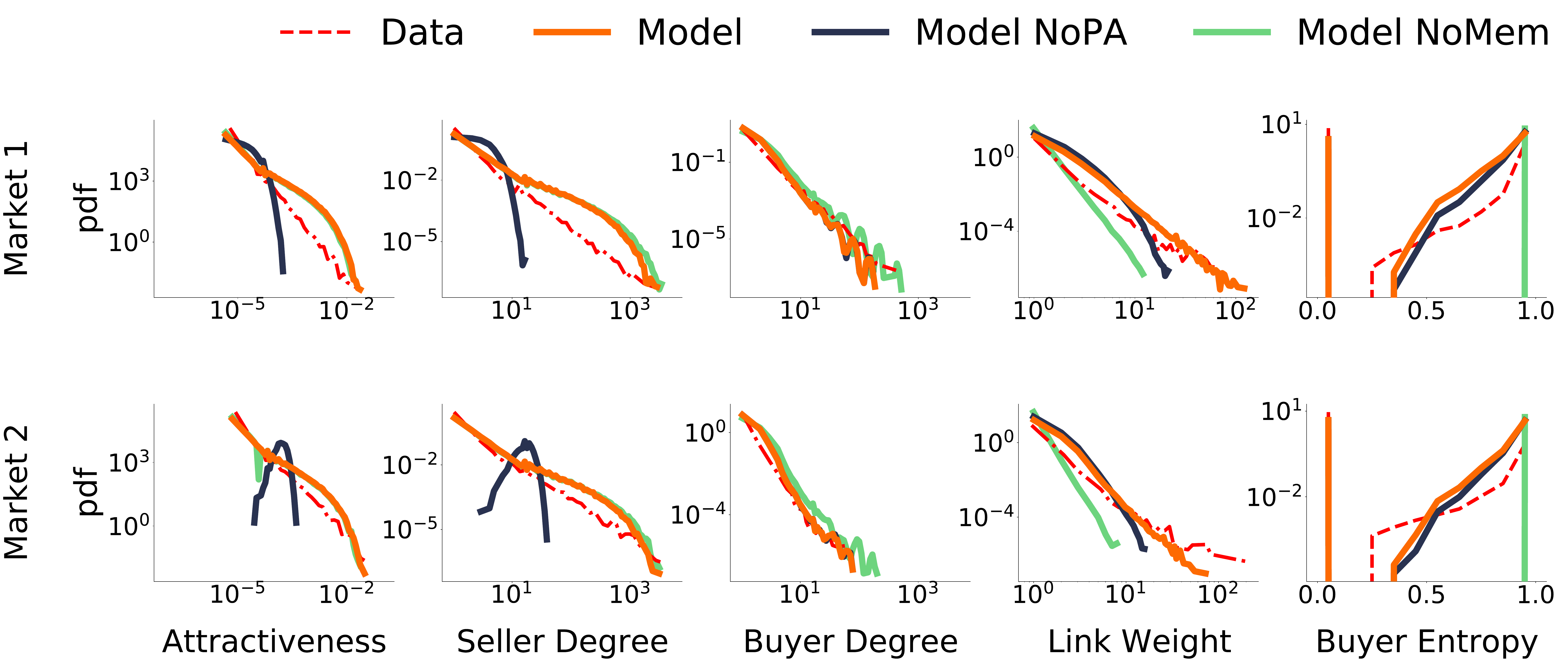}
	\caption{\label{fig:results_models} \textbf{The model reproduces different properties of buyer-seller networks.} 
	Each row corresponds to a different market (see Fig.~S2-S5 \SI for other markets), whose simulations parameters are individually calibrated as detailed in the main text. 
	From left to right, we show distributions for different quantities: attractiveness, seller degree, buyer degree, link weight and seller entropy. 
	The comparison with the two model variations, without preferential attachment or without memory, shows the key role of both parameters in shaping the network: preferential attachment is crucial in reproducing highly active sellers, whereas buyer memory is fundamental to capture the heterogeneity of buyer-seller relationships.}
\end{figure*}

Thus far, we have considered node-level properties aggregating detailed information on the links. For example, the attractiveness only accounted for the total number of links, whereas the degree only captures the total number of different buyers or sellers that the user has interacted with. 
To better understand how the model performs in reproducing finer details of the buyer-seller network structure, we test our model against two other properties of the aggregated network: link weight---the number of transactions between a buyer and a seller---and the buyer entropy, as defined in Eq.1. 
Our main model outperforms its two variations in reproducing the shape and tails of the link weight distribution. 
In particular, the memory mechanism appears to be fundamental in reproducing repeated transactions between a buyer and a seller.
The buyer entropy distribution is again well-captured by the model and shows how the memory mechanism is key to capturing the diversity of relationships buyers establish with different sellers. Indeed, the \textit{NoMem} model produces only entropy values close to 0 and 1; this happens because without memory,  a buyer almost never finds any previous seller, hence buyers making more than one purchase almost always buy from new sellers.

We have seen that our model is able to capture various aspects of the final aggregated buyer-seller network. 
The next step is to see whether our model can also reproduce the temporal evolution of the buyer-seller network. 
To investigate this, we focus on the degree of top sellers since we previously showed these sellers generate the largest activity and volume on these markets. 
We measure time by the total number of purchases made. 
Results are shown in Fig.5a-c, where we plot the temporal evolution of the top 50 (a), 100 (b) and 200 (c) sellers degree distribution for one illustrative product market. 
In doing so, we compare the model to its two variations and the data. 
Results for more product markets are shown in Fig.~S6-S9 \SI. 
The main model is able to reproduce the temporal evolution of the distributions, as clearly shown by the cores (i.e., interquartile ranges) overlapping at different times. 
We further compute the absolute value of the difference between the mean of the models' distribution and the mean of the data, for each of the nine equally spaced time steps and for all 28 simulated product markets. 
As shown in Fig.5d-f, the model is better able to reproduce the temporal dynamics for all simulated markets. 
Indeed, the median of the distance distributions is always smaller in the main model than the two other model variations.

\begin{figure*}[ht]
	\centering
	\includegraphics[width=16cm]{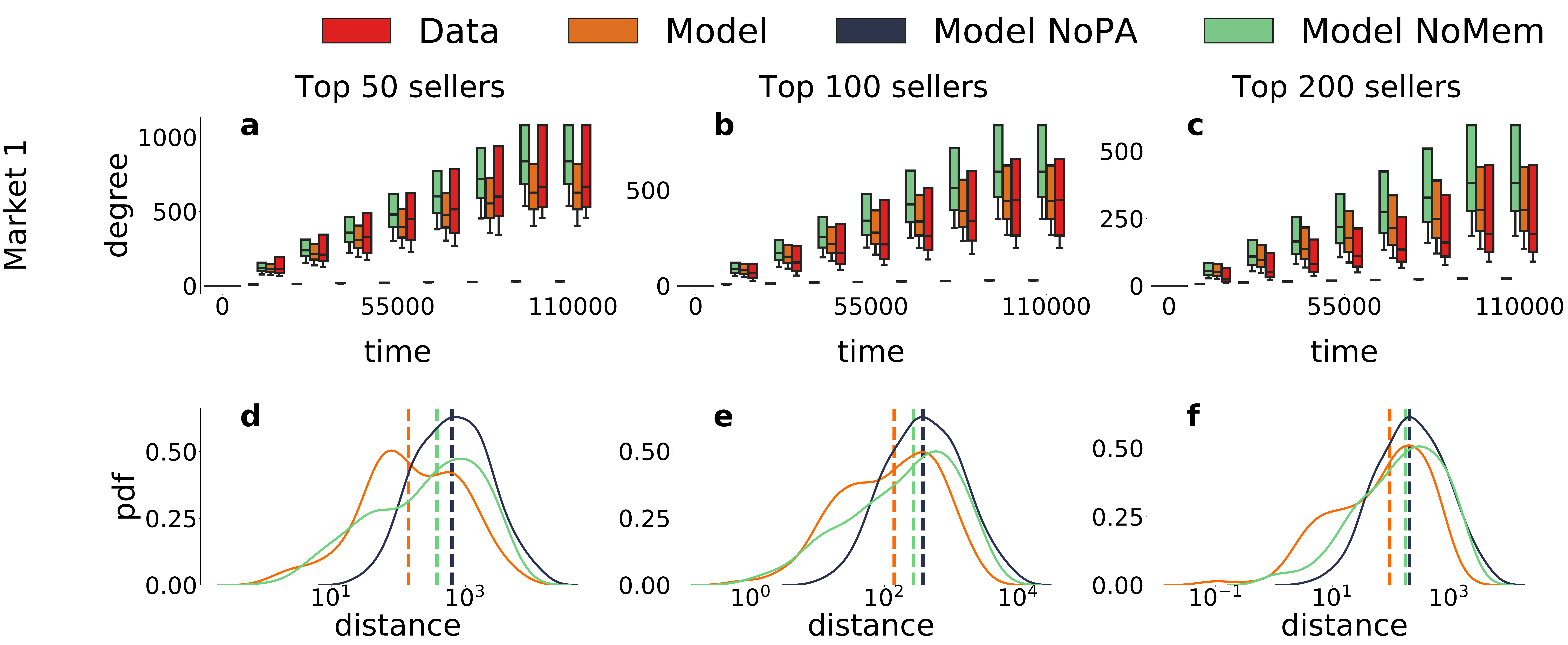}
	\caption{\label{fig:results_models_temporal} \textbf{Model reproduces the temporal evolution of the top sellers degree distribution.}
	Top (a to c): temporal evolution of the degree distribution of the top 50 (a), 100 (b) and 200(c) sellers, representing the distribution at 9 equally spaced time steps with boxplots ranging from the first to the third quartiles,  whiskers extending from $2.5^{th}$ to $97.5^{th}$ percentiles.
	Results are shown for one product market, all other markets are shown in Fig.~S6-S9 \SI. 
	Bottom (d to f): Distribution of the distance between the empirical and model(s) median degree of the top 50(left), 100(center) and 200(right) sellers, for all product markets and time steps, and the three considered models. 
	Vertical lines represent the distributions median, showing that the model median is always smaller than the alternatives.
	The model better captures the temporal evolution of the top sellers degree for all product markets than the alternatives neglecting either the preferential attachment or the memory mechanism.}
\end{figure*}

\section*{Discussion}
\label{sec:discussion}

In this paper, we have analyzed 244M (25B USD) transactions occurring on regulated and unregulated online marketplaces. 
First, we have revealed remarkable regularities in the aggregate static and temporal properties of the buyer-seller networks, both for buyers and sellers.
Then, we have revealed how buyers are affected by the memory of past interactions. 
Finally, we have proposed a model which captures the main stylized facts of the data, based only on three well-known network formation mechanisms of online marketplaces: buyers have different propensity to make purchases, they remember the sellers they purchased from, and they are more likely to buy from successful sellers.

It is important to highlight the limitations of our study, which also represent directions for further work.
First, while our study is based on (pre-processed) blockchain data, 
access to DWM server logs could provide more detailed information on some specific markets, for instance, the directed buyer-seller links which are not observable in our data. 
Second, 
the model could be further enriched with other known mechanisms: pricing dynamics~\cite{goolsbee2018internet,gorodnichenko2018price}, product search ranking~\cite{consumer_search_1,consumer_search_2,consumer_search_3,consumer_search_4}, customer reviews~\cite{negative_reputation}, sellers entering or leaving the platform~\cite{nosko2015limits}, and recommendation algorithms~\cite{recommender_1}. 
Finally, including richer economic incentives (e.g., strategic behavior) to model buyers' and sellers' decisions could shed light on how agents could exploit their market power.
A deeper understanding of economic incentives and equilibrium behavior in buyer-seller networks could ultimately inform market design and regulation of online marketplaces.

Nevertheless, our work supports and extends previous findings. 
The fat-tailed heterogeneous curves in Fig.1a-d substantiate previous observations of high concentration in DWMs: wholesale~\cite{geography_1}, few sellers~\cite{closure_1} or few buyers~\cite{specific_2} were found responsible for the largest part of volumes in smaller samples of data. 
The fat-tailed inter-event time distributions, spanning times between a second and a year, are compatible with the bursty nature of several social activities~\cite{burstiness_1,burstiness_3}, and the finding about a shared memory kernel further points to a similarity between social and economic activities~\cite{karsai_time_2015,ubaldi_asymptotic_2016}. 
Taken together, our results could inform and enrich economic models where heterogeneity assumptions are now commonplace~\cite{competition_2} but empirical evidence on the structure of buyer-seller networks has not yet been introduced.

The regularities observed in Fig.1 are surprising given
the differences in the marketplaces covered by our data: transactions on the clear web with state enforcement of contracts~\cite{regulated_vs_unregulated} vs. transactions on the dark web that rely mainly on reputation and self-governance~\cite{reputation_dwm_1}; the sale of only regulated products vs. mainly unregulated products; the use of fiat vs. the use of cryptocurrencies. 
And, indeed, there is both substantial heterogeneity in product markets in the e-commerce dataset and several differences across marketplaces in the DWM dataset (e.g., existence time period, geography, product focus, etc.).
Our model suggests specific mechanisms that drive the regularities across the two datasets.
Sellers build a reputation that makes them more attractive to buyers who, in turn, are affected by their memory of the sellers they already purchased from. 
In particular, the presence of both memory and preferential attachment is fundamental in reproducing both local and global properties of the buyer-seller network, as already shown for the intrinsically different social networks~\cite{social_systems_1,social_systems_3,burstiness_1,karsai_time_2015}. 
However, commercial interactions exhibit important differences compared to social interactions, with preferential attachment playing a dominant role in the market dynamics.

Our results point towards alternative strategies to attempt to reduce trading of illicit goods on dark web marketplaces. 
Historically, DWMs have been closed after long and expensive operations targeting the market admins in order to arrest them and shut down the servers~\cite{europol_report_1}. 
However, the high degree of concentration, the importance of preferential attachment, and the memory kernel in the buyer dynamics, all suggest that limited observations of the market dynamics could give a clear enough picture of who the key actors of these networks are. 
For instance, key sellers will most likely attract most of the observed purchases from the more active buyers, and stopping them would effectively stop a large part of the market trade. 
In this regard, our model could also be used to produce candidate synthetic DWM buyer-seller networks to quantitatively study and simulate the effects of targeting ``key players'' on marketplaces~\cite{ballesterWhoWhoNetworks2006}.

Finally, a better understanding of buyer-seller network formation could have consequences for market design and regulation. 
For example, fat-tailed distributions show a high degree of concentration on both buyer and seller sides of the marketplaces: just a few agents (both on the buyers and seller sides) are responsible for a vast majority of the transaction volume.
While buyer market power appeared in analyzes of labor monopsony and retailers~\cite{buyer_power_better_1,buyer_power_better_2}, our empirical finding of buyer concentration calls for a deeper understanding of buyer power in online marketplaces. 
   Moreover, these observations can also inform theoretical economic models of online marketplaces, providing empirical backing to heterogeneity assumptions and suggesting specific values for parameters or shapes for distributions.  
Also, we find signs of both local (memory) and global (reputation) mechanisms in the structure and evolution of buyer-seller relationships. 
Thus, the inclusion of memory and reputation in previously developed models can improve our understanding of the pricing of network effects~\cite{market_power_better_2}, inter-platform competition~\cite{competition_2} and long-run sustainability of the platforms~\cite{sustainability_better_1}.

\section*{Materials and Methods}

\subsection*{Dark Web Marketplaces}

Dark Web Marketplaces are illegal unregulated commercial websites.
They operate similarly to other online marketplaces, such as Gumtree or Craigslist. 
To improve anonymity DWMs are reached through browsers supporting the onion protocol, and use cryptocurrencies, mainly Bitcoin, as the main currency. 
While all Bitcoin transactions are publicly available, they record money exchanges between addresses, and a user can generate a new address (identifier) at each transaction to favour anonymity. 
As a result, the data need to be pre-processed to cluster addresses into individual entities in order to perform any economic analysis. 
Our dataset has been pre-processed by Chainalysis Inc.~\cite{chainalysis} to map addresses into entities (see \textit{Supplementary Information} for more details). 

Our dataset contains the entire transaction data of 28 entities corresponding to DWMs between June, 2011, and February 2021 (see Fig.~S10 and Tab.~S2 \SI). 
These markets all have an average daily volume of more than 15,000 USD, in order to be able to reliably measure different observables, and include all relevant DWMs as identified by law enforcement agencies~\cite{europol_report_1}. 
The data contain all transactions received or sent by DWMs, excluding services such as exchanges (Bitcoin trading exchanges allow users to trade Bitcoin). 
Note that the data hide the direct buyer-seller link, because the money pass through the platform during the transaction. 

\subsection*{E-Commerce Platform}

E-commerce platforms are regulated online marketplaces where sellers can post ads for products. 
Buyers and sellers can generally be both individuals or businesses. 
The payment is usually processed by the platform, but the shipping is handled by the seller. 
Sellers receive feedback from buyers, which together with product categorization helps people navigate the platform and choose what to buy. 

The data used in this study consist in all the purchases made on 144 product markets from a popular e-commerce platform since 2010. The 144 product markets have been randomly selected from the markets that were active during the entire time period. The data cover only one geographical region. 
Similarly to the DWM data, the transaction data include: timestamp of the transaction, pseudonyms for buyer and seller, and the amount spent in the transaction. 
One key difference is that the data show the direct link between buyer and seller, forming a bipartite buyer-seller network and allowing for a more fine-grained analysis. 
For more details on the data see Fig.~S11 \SI. 

\subsection*{Model simulation} \label{model_simulations}

Each simulation is tuned to simulate one specific product market.
We fix the agents population according to the data: number of sellers $N$, number of buyers $M$, and simulation total number of time steps $T$, to fix the average total number of transactions in our simulations as in the data:  $\langle a_i  \cdot \Delta t \cdot T \cdot N \rangle = t$, where $a_i$ is the buyer activity as defined in the main text, $\Delta t$ is the simulation time step (fixed to 1) and $t$ is the total number of transactions present in the data. 
We realise 30 different realizations for each parameter set, and aggregate the final results.

\section*{Acknowledgments} 
A.Br. and A.Ba. gratefully acknowledge support from the Alan Turing Institute. J.B. gratefully acknowledges the support of the Center of Mathematical Sciences and Applications at Harvard University.
\section*{Author Contributions} 
All authors designed the research; A.Br., J.B. and A.E. acquired the data. A.Br. and J.B. performed the measurements. 
All authors analyzed the data. All authors wrote the manuscript, discussed the results and commented on the manuscript.\\
\section*{Competing Interests} 
The authors declare that they have no competing interests.\\
\section*{Data and materials availability} 
All data needed to evaluate the conclusions in the paper are present in the paper and/or the Supplementary Materials. 
Additional data related to this paper may be requested from the authors.

% \clearpage
% \bibliographystyle{unsrt}
% \bibliography{bibtex_online_marketplaces}

\clearpage
\onehalfspacing
\begin{Large}{\begin{center}
{\bf{Supplementary Information for Macroscopic properties of buyer-seller networks in online marketplaces}}
\end{center} 
 }
\end{Large}
\thispagestyle{empty}
\tableofcontents
\clearpage

\appendix

\renewcommand{\figurename}{Supplementary Figure}

\setcounter{figure}{0}

\renewcommand{\tablename}{Supplementary Table}

\section{Supplementary Text}
\subsection{Dark Web Marketplaces}

Dark Web Marketplaces (DWMs) are unregulated commercial websites. 
They operate similarly to other online marketplaces, such as Gumtree or Craigslist.
In most cases, the buyer sends money to the DWM which keeps the money until the client has confirmed receipt of the goods, upon which the funds are released to the seller after taking a small fee. 
Customers may also leave reviews, contributing to the sellers reputation. 
To favour anonymity DWMs are reached through browsers supporting the onion protocol~\cite{tor}, to find their updated domain one can navigate one of the many specialised websites like DeepDotWeb and darknetlive~\cite{DarkNetLive}. 
DWMs use cryptocurrencies, mainly Bitcoin, as the main currency. 

Our dataset contains the entire transaction data of 28 dark marketplaces between June, 2011, and July 2020. 
These markets all have an average daily volume larger than 15,000 USD, in order to be able to reliably measure different observables, and include all relevant DWMs as identified by law enforcement agencies. 
Each marketplace can be represented as an egocentric network around the DWM, a star where the marketplace is the central node and its nearest neighbours represent marketplace users. 
A directed edge represents a transaction occurring between the DWM and one of its nearest neighbours. 
Note that the data hide the buyer-seller direct link, because the money go through the platform during the purchase. 

All transactions ever done in Bitcoin are publicly available, and can be downloaded by installing Bitcoin core software or through various third party APIs like Blockchain.com. 
Each transaction is recorded through its time, exchanged amount, source and destination addresses. 
An address is an alphanumeric identifier, and a user can generate a new one every time he does a transaction, and this what now generally happens for anonymity reasons. 
For this reason, the data need to be pre-processed to cluster addresses into individual entities in order to perform any economic analysis. 

Our dataset has been pre-processed by Chainalysis Inc following approaches as in ~\cite{btc_stateart_1,btc_stateart_2,btc_stateart_3}. 
This process uses established heuristics~\cite{btc_heuristic_1,btc_heuristic_2,btc_heuristic_3,btc_heuristic_4,btc_heuristic_5} in order to map addresses into entities. 
This process is unsupervised, as there is no ground truth data to rely on. 
Avoiding false positives in the clustering is crucial, as this data have been used in many law enforcement investigations related to illicit activities~\cite{chainalysis_news_1}. 
This means that if an addresses does not meet the heuristics conditions it is left unclustered, and therefore not all addresses linked to DWMs or their admins may be correctly included in the market entity. 

Chainalisis Inc. also identifies the clustered entities corresponding to a number of dark web marketplaces. 
Our dataset contains their full transaction history, meaning all the transaction in which the markets either send or receive money from other entities.
Bitcoin trading exchanges were excluded from the list of trading entities, since we focus on the users’ direct interaction with the dark marketplace. 
(Bitcoin trading exchanges are platforms that allow users to trade Bitcoin for other cryptocurrencies or fiat currencies.)

We collect additional information on the analysed marketplaces from different sources, including the Gwern archive~\cite{gwern}, law enforcement agencies reports and dedicated online forums. 
We focused our attention on the creation and closure dates of these markets, in order to correctly interpret the transaction data. 
We report the lifetimes of the selected markets in Fig.~\ref{fig:life_markets}, color coding by the daily average number of transactions as proxy for the market size. 
For more details on the markets, see also Tab.~\ref{tab:dwm_numbers}.

\subsection{Computing the Memory Kernel}

In this section we describe the computations made in order to estimate the buyer memory parameters $c$,$\beta$. 
As detailed in the main text, buyers are grouped in different classes according to their final degree (number of different sellers they purchased from) at the end of the considered periods. 
Classes are divided in powers of 2, for example the first class includes buyers with degree 1, the second with degree between 2 and 3, the third between 4 and 7 and so on. 
In our dataset, a buyer has a unique identifier in each market product, not allowing to follow their behavior across different markets. 
In order to reduce the noise in the data, all markets are aggregated together for the following computations.

In order to estimate the memory parameters from eq.2, defined in the main text, the first step is to estimate the conditional probabilities $P_k(n+1|n) = P_k(n)$ of buying from a new $n+1^{th}$ seller when you already bought from $n$ different ones. 
To do so, we count the number of buyers $b_k(n)$  in class $k$ who go from degree $n$ to $n+1$, and we count the total number of purchases $p_k(n)$ they made when they had degree $n$.

\begin{equation}
    P_k(n) = b_k(n) / p_k(n)
\end{equation}

In order to reduce the noise on the computation of $P_k(n)$, we limit the computation to $n \leq k$. 
This way, all buyers in degree class $k$ go from degree $n$ to $n+1$, as their final degree is at least equal to $k$.
The numerator is therefore constant, and equal to $N_k$, the number of buyers in degree class $k$.
Equation $P(n)$ then reads:

\begin{equation}
    P_k(n) = N_k / p_k(n)
\end{equation}

Assuming that for a given degree $n$ events are independent, or in other words that users behave independently of each other, and checking that $1 << N_k << e_k(n)$, we can estimate the uncertainty of $P_k(n)$ as follows:
\begin{equation}
    \sigma(P_k(n)) = \sigma_k(n) = \sqrt{P_k(n)(1 - P_k(n)) / e_k(n)}
\end{equation}
Having estimated the curve $P_k(n)$ for each degree class $k$, we can fit eq.2 to each curve separately.
To do so, we do a numerical least square optimization, estimating the values of $\beta$ and $c$ for each degree class. 
Results are shown in Tab.~\ref{tab:fit_coefficients}.

\subsection{Sampling of Product Markets}
The e-commerce platform contains data on 144 product markets. 
We sample 28 DWMs to fit with our model. 
The 28 product markets are sampled to ensure all products are represented.
In particular, products can be grouped together in higher-level markets, from which we sample on product each.
To make an example: our dataset may contain two product markets in the fruit group, namely apples and pears.
In the sample for the model simulation we choose only one of the two, taking care that the 28 sampled product markets are representative of the heterogeneous market size of our dataset.

\subsection{Maximum Likelihood Estimation}
As detailed in the main text, we employ a data-driven approach to estimate the model parameters for each product market. 
In particular, the preferential attachment parameter $\mu$, describing the increment of a seller's attractiveness after a sale, is estimated by Maximum Likelihood Estimation (MLE). 
To do so, we simulate the model for each value of $\mu$ on a grid, ranging from 1 to 500, and then compute the associate negative log-likelihood computed comparing the data to the simulated attractiveness distribution, and choose the value minimizing the quantity. 
We employ this simple approach to estimate $\mu$, by only analysing the attractiveness distribution, as the scope of this work is to study and reproduce stylized facts, and not to propose a detailed model precisely reproducing all details of a given product market. 
For this reason, even the value of $\mu$ itself assumes relative importance, as its order of magnitude determines the agreement with the data, but small variations in the precise value are meaningless in the context of this study. 
For completeness, in Tab.~\ref{tab:delta} we show the fitted value of $\mu$ for each product market.

\section{Supplementary Figures}

\begin{figure*}[ht]
	\centering

	\includegraphics[width=\figureSmaller]{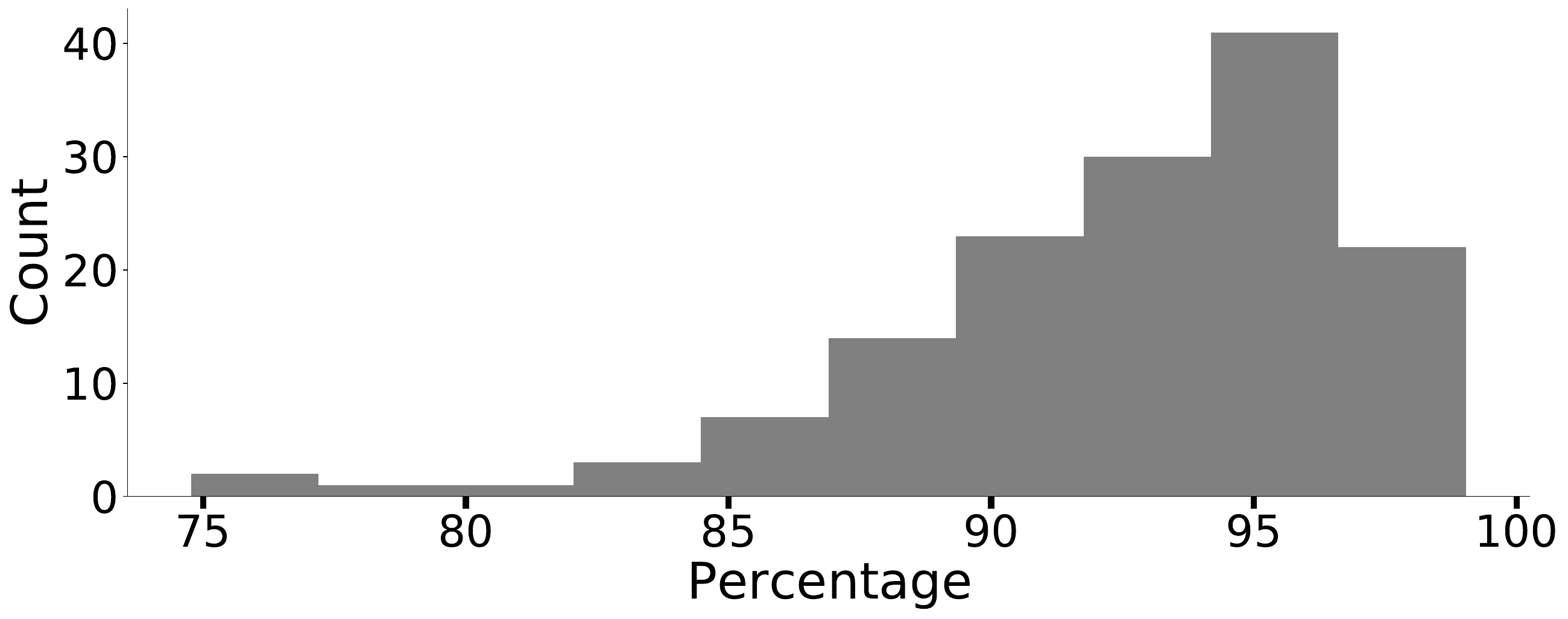}
	\caption{\label{fig:one_trx}\textbf{Most buyers with zero entropy only have done one transaction.} Histogram of percentage of buyers doing just one transaction for each e-commerce product market, among those with zero entropy. In most markets the percentage is well above 90\%.  }

\end{figure*}

In Fig.~\ref{fig:one_trx}, we show the histogram of the percentage of users with entropy zero doing just one transaction, in each product market. 
This percentage is always greater than $75\%$, but actually over $90\%$ in most cases, showing how buyers with entropy zero can effectively be neglected when showing the buyer entropy distribution.

\begin{figure*}[ht]
	\centering

	\includegraphics[width=\figureDefaultWidth]{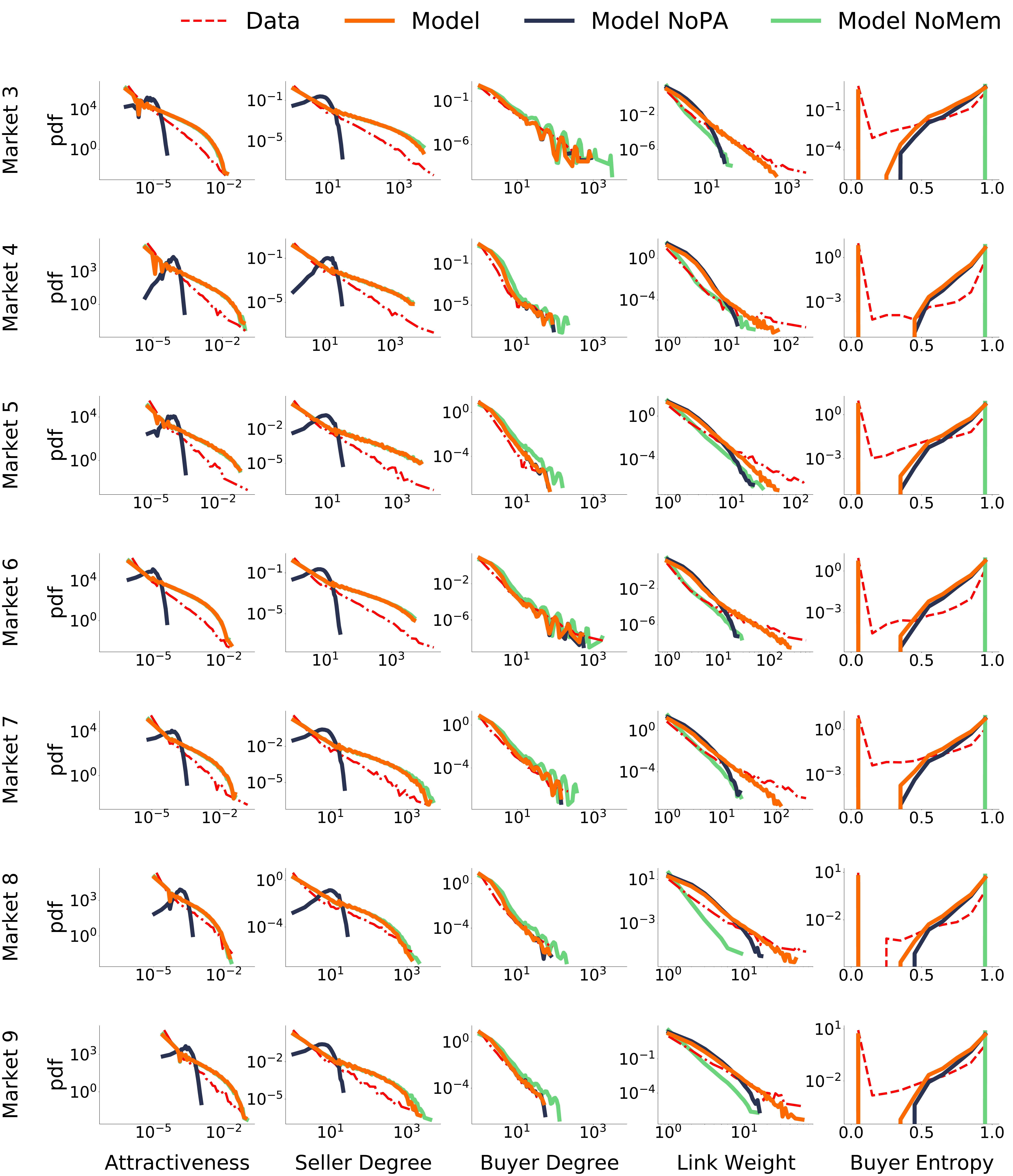}
	\caption{\label{fig:other_cats_static} \textbf{Model simulations for different markets - final distributions - Markets 3 to 9.} 
	Each row corresponds to a different market, whose simulations parameters are individually calibrated as detailed in the main text. 
	From left to right, we show distributions for different quantities: attractiveness, seller degree, buyer degree, link weight and seller entropy. 
	The comparison with the two model variations, without preferential attachment or without memory, shows the key role of both parameters in shaping the network: preferential attachment is crucial in reproducing highly active sellers, whereas buyer memory is fundamental to capture the heterogeneity of buyer-seller relationships.}

\end{figure*}

\begin{figure*}[ht]
	\centering

	\includegraphics[width=\figureDefaultWidth]{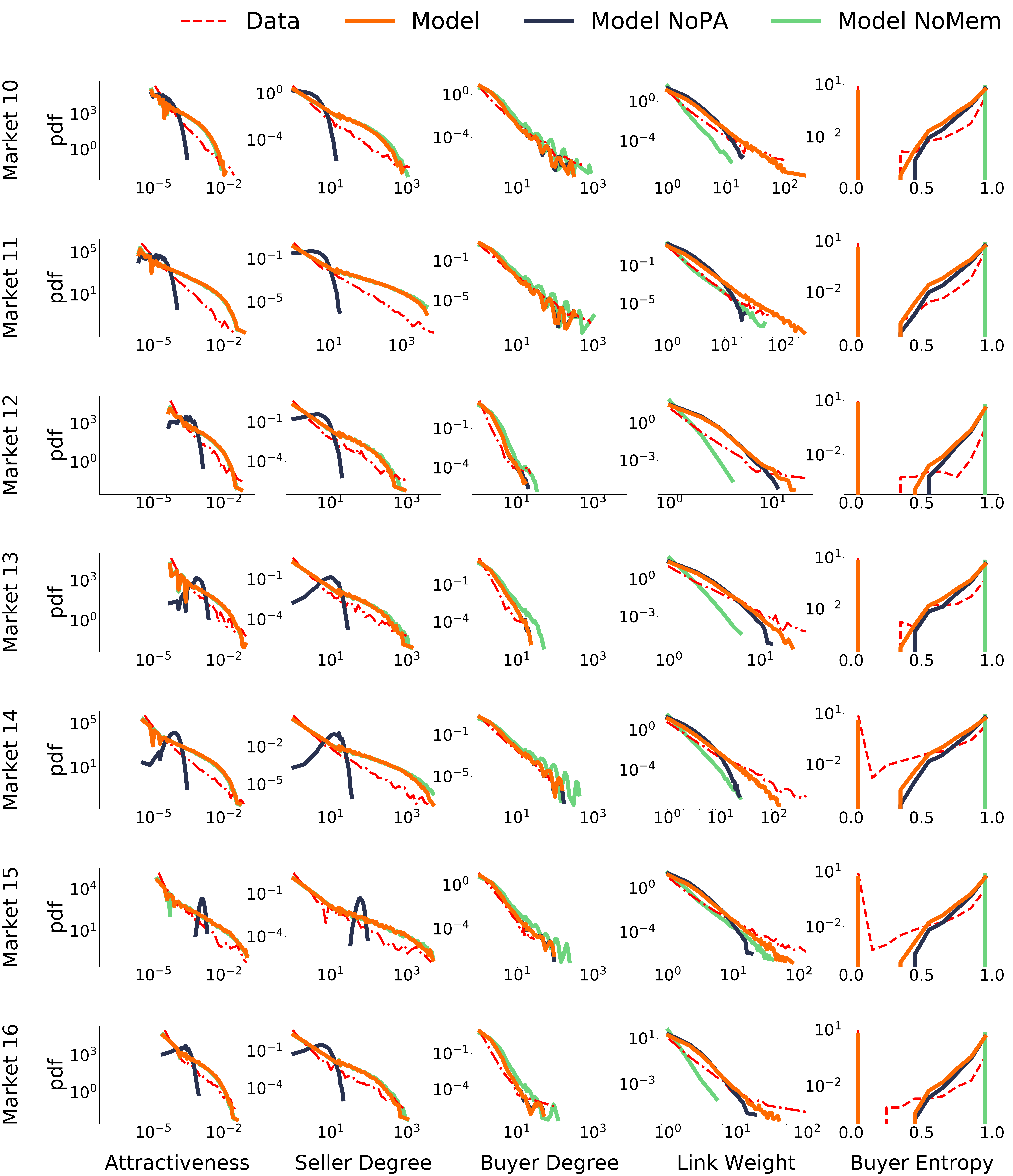}
	\caption{\label{fig:other_cats_static2} \textbf{Model simulations for different markets - final distributions - Markets 10 to 16.} 
	Each row corresponds to a different market, whose simulations parameters are individually calibrated as detailed in the main text. 
	From left to right, we show distributions for different quantities: attractiveness, seller degree, buyer degree, link weight and seller entropy. 
	The comparison with the two model variations, without preferential attachment or without memory, shows the key role of both parameters in shaping the network: preferential attachment is crucial in reproducing highly active sellers, whereas buyer memory is fundamental to capture the heterogeneity of buyer-seller relationships.}

\end{figure*}

\begin{figure*}[ht]
	\centering

	\includegraphics[width=\figureDefaultWidth]{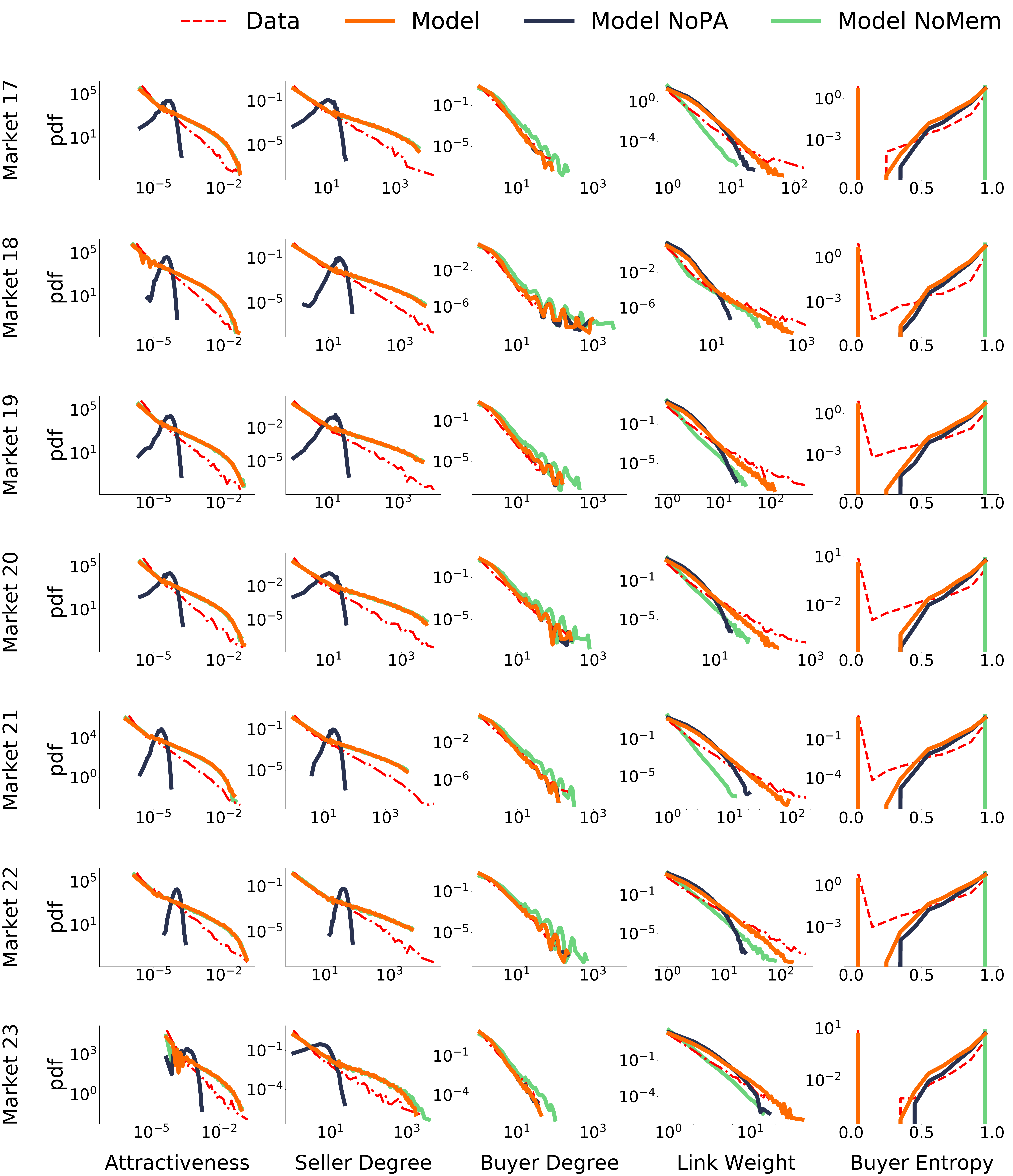}
	\caption{\label{fig:other_cats_static3} \textbf{Model simulations for different markets - final distributions - Markets 17 to 23}
	Each row corresponds to a different market, whose simulations parameters are individually calibrated as detailed in the main text. 
	From left to right, we show distributions for different quantities: attractiveness, seller degree, buyer degree, link weight and seller entropy. 
	The comparison with the two model variations, without preferential attachment or without memory, shows the key role of both parameters in shaping the network: preferential attachment is crucial in reproducing highly active sellers, whereas buyer memory is fundamental to capture the heterogeneity of buyer-seller relationships.}

\end{figure*}

\begin{figure*}[ht]
	\centering

	\includegraphics[width=\figureDefaultWidth]{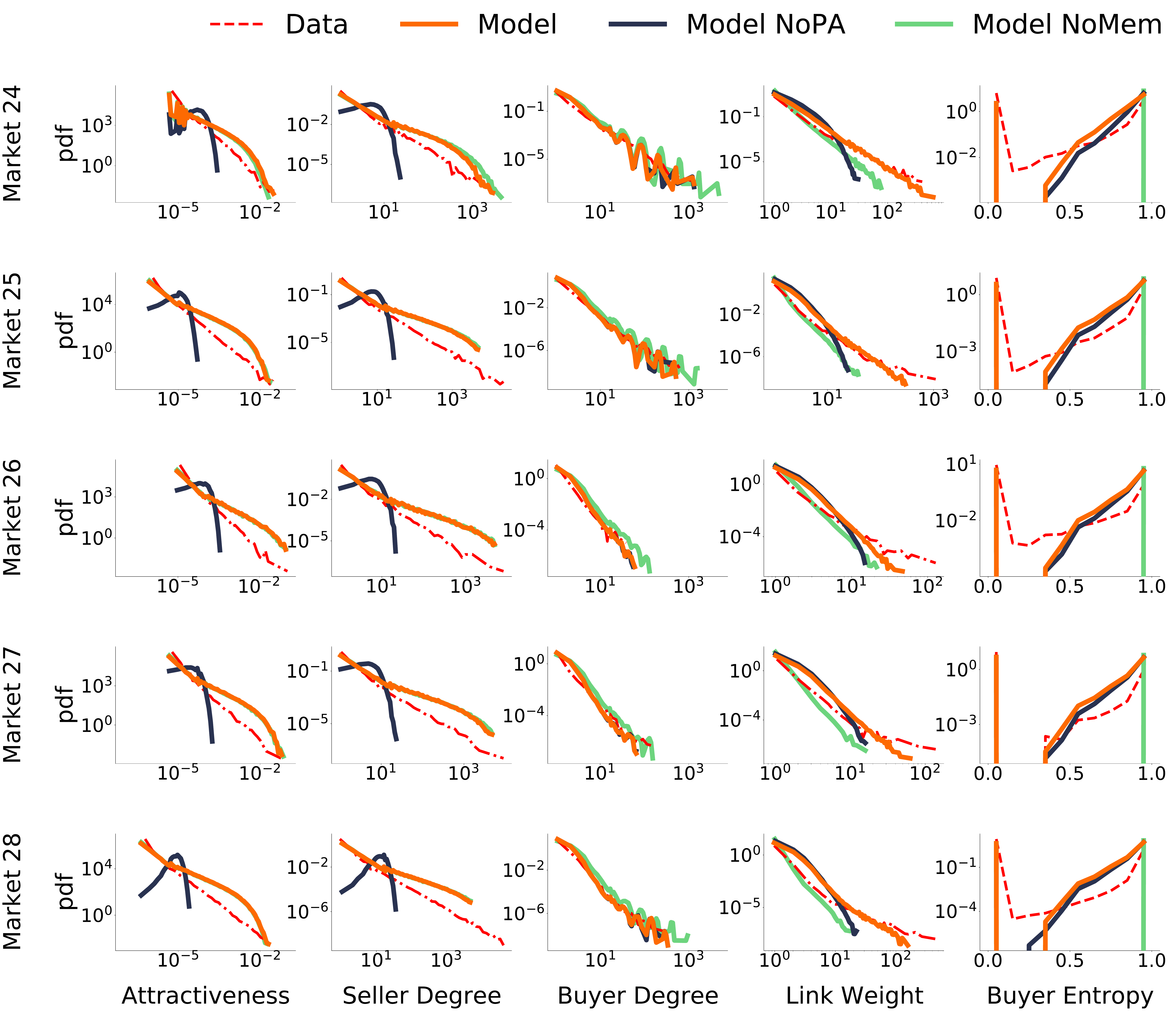}
	\caption{\label{fig:other_cats_static4} \textbf{Model simulations for different markets - final distributions - Markets 24 to 28}
	Each row corresponds to a different market, whose simulations parameters are individually calibrated as detailed in the main text. 
	From left to right, we show distributions for different quantities: attractiveness, seller degree, buyer degree, link weight and seller entropy. 
	The comparison with the two model variations, without preferential attachment or without memory, shows the key role of both parameters in shaping the network: preferential attachment is crucial in reproducing highly active sellers, whereas buyer memory is fundamental to capture the heterogeneity of buyer-seller relationships.}

\end{figure*}

In Fig.~\ref{fig:other_cats_static} to ~\ref{fig:other_cats_static4} we show results of model simulation for 26 other product markets. 
The results show how the model is able to capture the main stylised facts of the buyer-seller network structure, with memory and preferential attachment both necessary to capture different aspects of the structure.

\begin{figure*}[ht]
	\centering

	\includegraphics[width=\figureDefaultWidth]{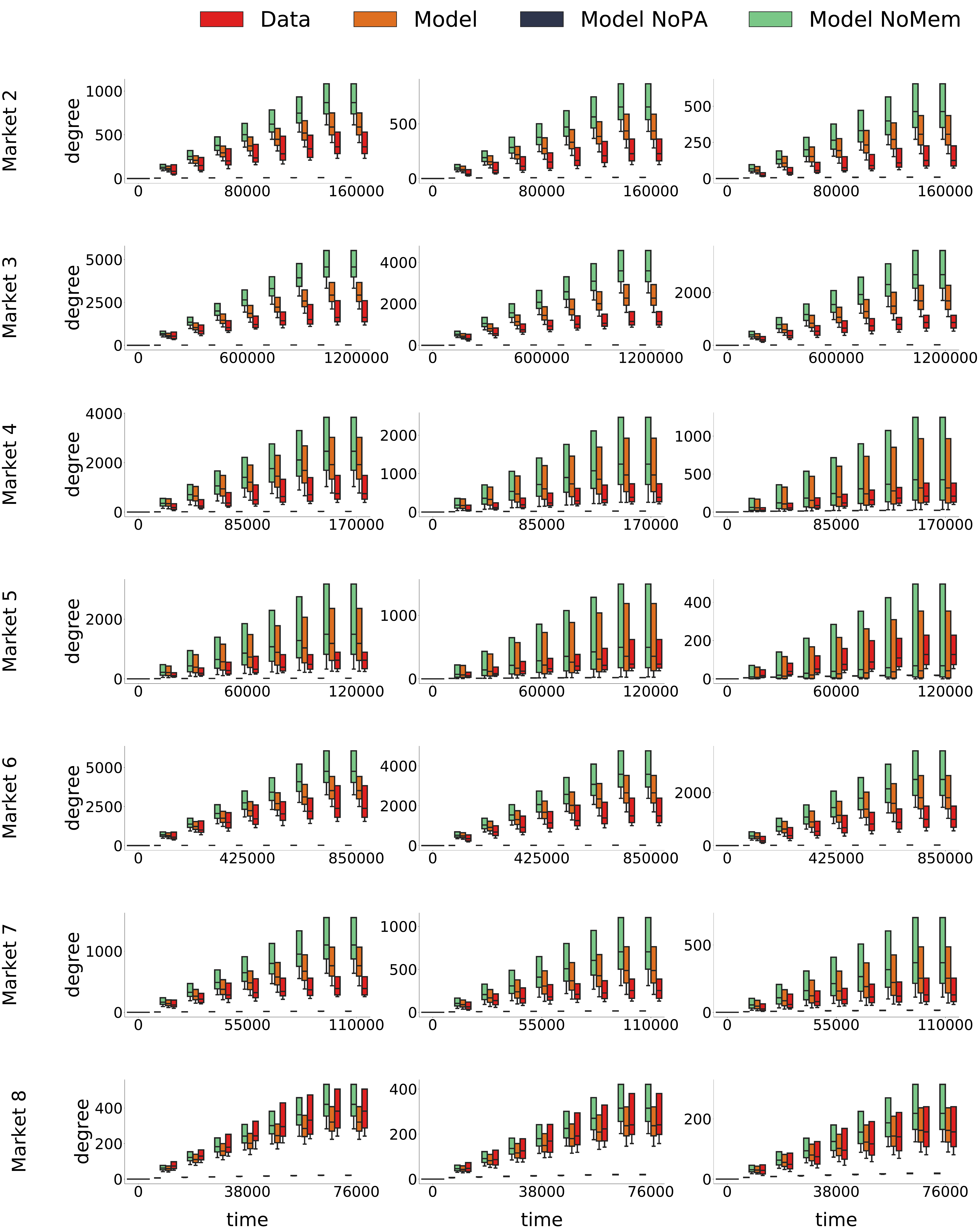}
	\caption{\label{fig:other_cats_temporal} \textbf{Model simulations for different markets - temporal evolution - Markets 2 to 8} 
	Each row represents one market.
	From left to right: temporal evolution of the degree distribution of the top 50 (left), 100 (center) and 200(right) sellers, representing the distribution at 9 equally spaced time steps with boxplots ranging from the first to the third quartiles,  whiskers extending from $2.5^{th}$ to $97.5^{th}$ percentiles. 
	The model better captures the temporal evolution of the top sellers degree for all product markets than the alternatives neglecting either the preferential attachment or the memory mechanism.}

\end{figure*}

\begin{figure*}[ht]
	\centering

	\includegraphics[width=\figureDefaultWidth]{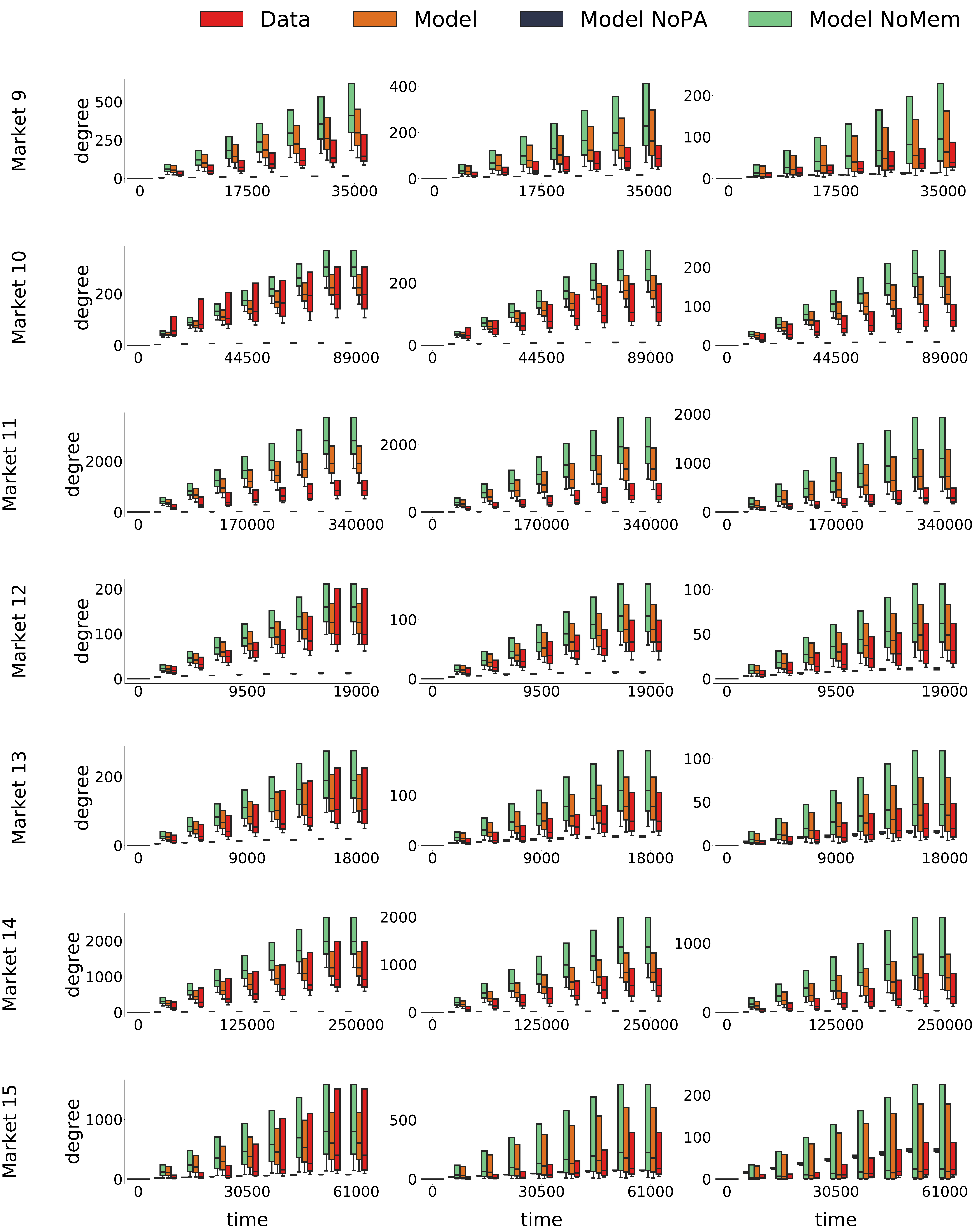}
	\caption{\label{fig:other_cats_temporal2} \textbf{Model simulations for different markets - temporal evolution - Markets 9 to 15} 
	Each row represents one market.
	From left to right: temporal evolution of the degree distribution of the top 50 (left), 100 (center) and 200(right) sellers, representing the distribution at 9 equally spaced time steps with boxplots ranging from the first to the third quartiles,  whiskers extending from $2.5^{th}$ to $97.5^{th}$ percentiles. 
	The model better captures the temporal evolution of the top sellers degree for all product markets than the alternatives neglecting either the preferential attachment or the memory mechanism.}

\end{figure*}

\begin{figure*}[ht]
	\centering

	\includegraphics[width=\figureDefaultWidth]{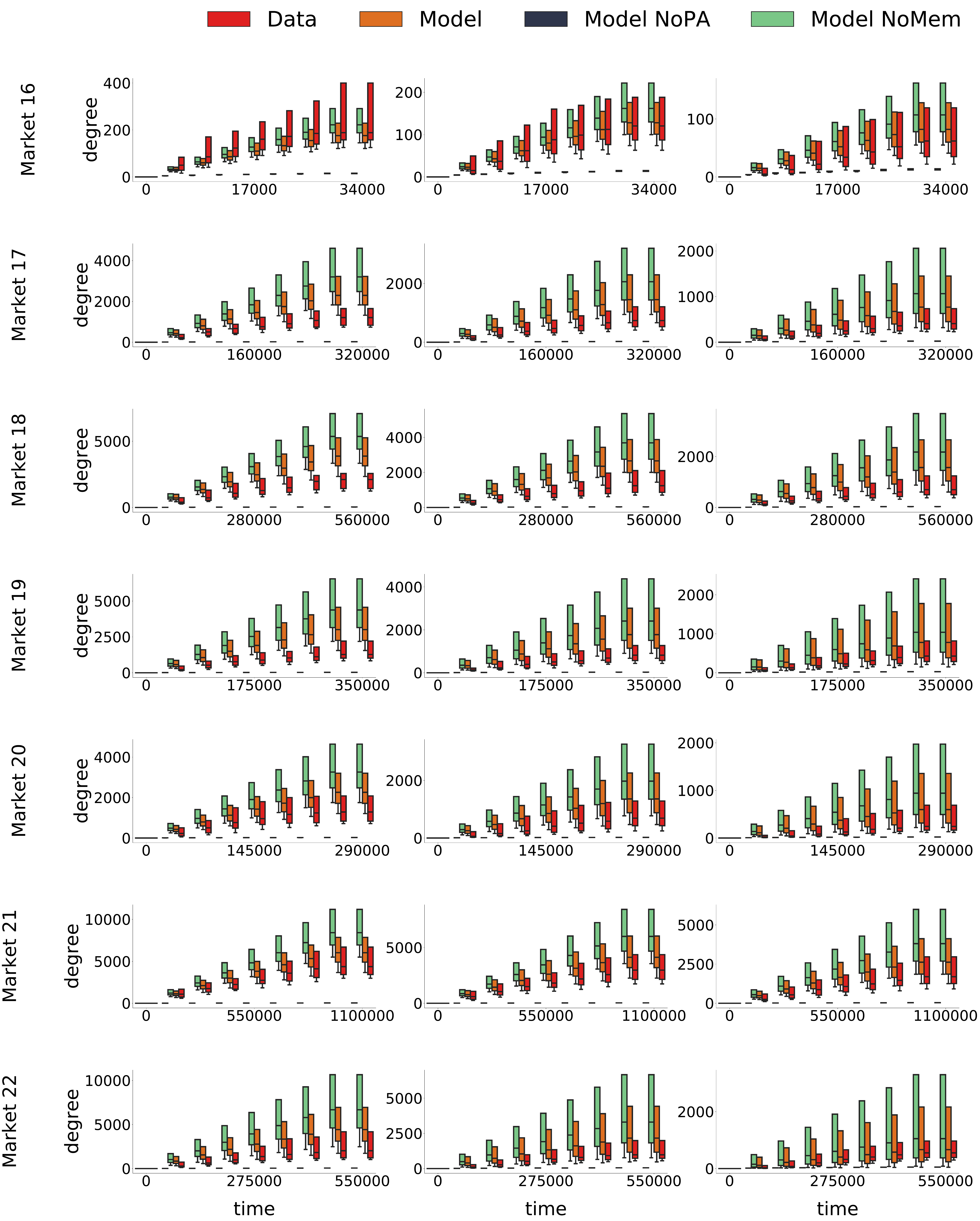}
	\caption{\label{fig:other_cats_temporal3} \textbf{Model simulations for different markets - temporal evolution - Markets 16 to 22} 
	Each row represents one market.
	From left to right: temporal evolution of the degree distribution of the top 50 (left), 100 (center) and 200(right) sellers, representing the distribution at 9 equally spaced time steps with boxplots ranging from the first to the third quartiles,  whiskers extending from $2.5^{th}$ to $97.5^{th}$ percentiles. 
	The model better captures the temporal evolution of the top sellers degree for all product markets than the alternatives neglecting either the preferential attachment or the memory mechanism.}

\end{figure*}

\begin{figure*}[ht]
	\centering

	\includegraphics[width=\figureDefaultWidth]{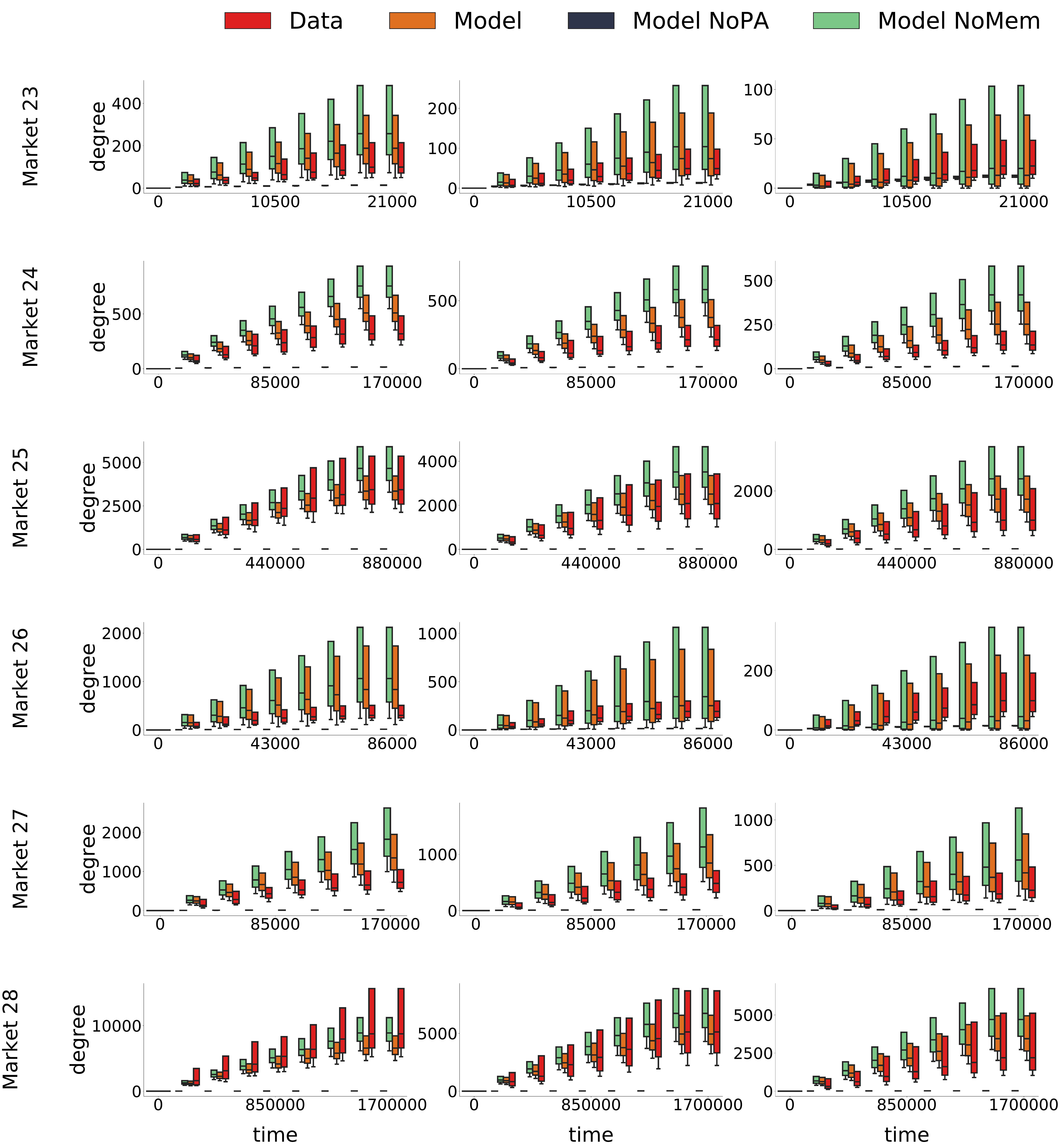}
	\caption{\label{fig:other_cats_temporal4} \textbf{Model simulations for different markets - temporal evolution - Markets 23 to 28} 
	Each row represents one market.
	From left to right: temporal evolution of the degree distribution of the top 50 (left), 100 (center) and 200(right) sellers, representing the distribution at 9 equally spaced time steps with boxplots ranging from the first to the third quartiles,  whiskers extending from $2.5^{th}$ to $97.5^{th}$ percentiles. 
	The model better captures the temporal evolution of the top sellers degree for all product markets than the alternatives neglecting either the preferential attachment or the memory mechanism.}

\end{figure*}

In Fig.~\ref{fig:other_cats_temporal} to ~\ref{fig:other_cats_temporal4} we show the temporal evolution of the top 50, 100 and 200 sellers degree distribution for 26 other product markets, represented as boxplot for 9 equally spaced time steps. 
The model is consistently able to reproduce the temporal evolution of the degree distribution, as shown by the cores of the boxplots (interquartile range) overlapping.

\begin{figure*}[ht]
	\centering

	\includegraphics[width=\figureDefaultWidth]{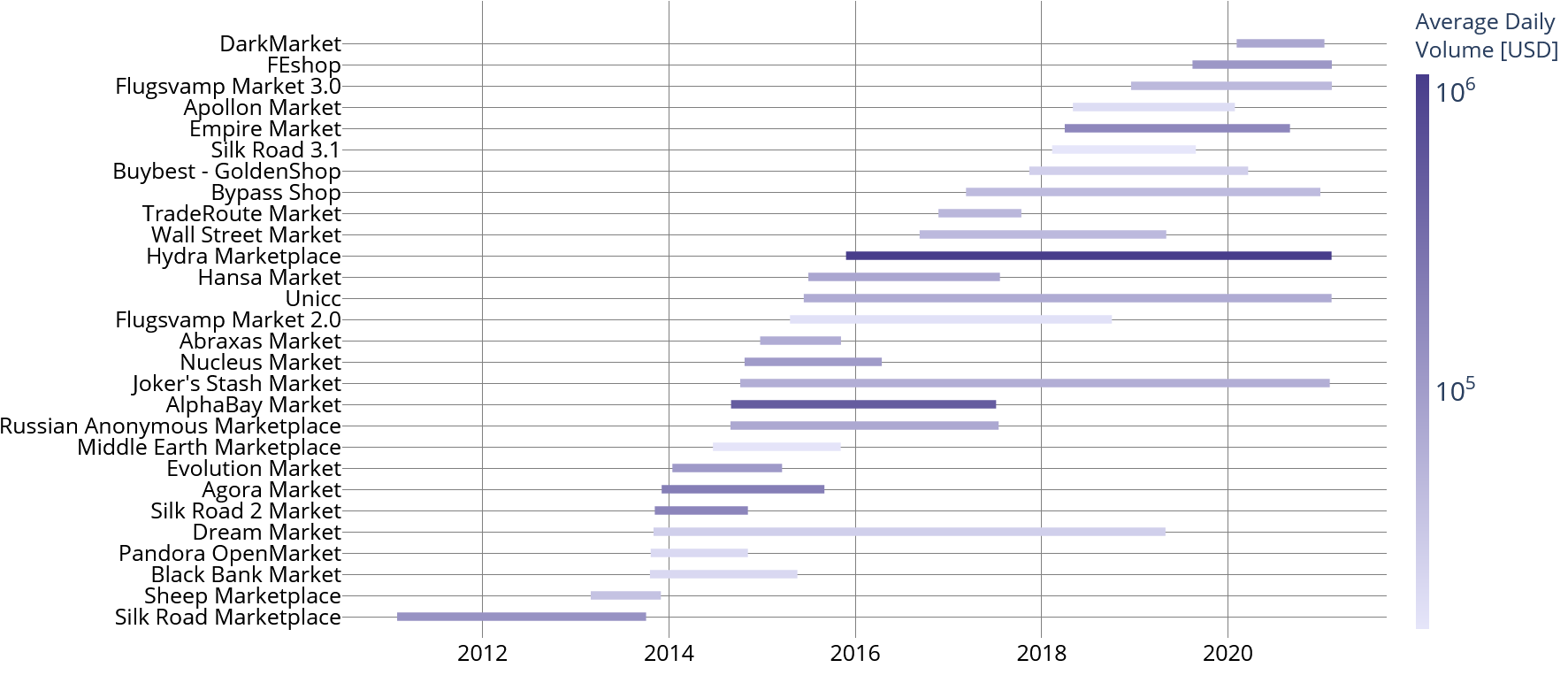}
	\caption{\label{fig:life_markets}\textbf{Dark Web Marketplaces} Duration of each market in our dataset, color coded by the number of transactions involving each marketplace. 
	Each market is live at least for 180 days and averages at least 20'000 transactions per day.}

\end{figure*}

In Fig.~\ref{fig:life_markets} we show the duration of each DWM in our dataset, color-coding by the average daily volume of transactions in USD. 
Our dataset covers all major DWMs from their onstart in 2011 with Silk Road Marketplace. 
DWMs are heterogeneous in daily volume, with some being just over our threshold of 20,000 USD, and Hydra Marketplace or AlphaBay Market close to 1M daily USD.

\begin{figure*}[ht]
	\centering
	\includegraphics[width=\figureDefaultWidth]{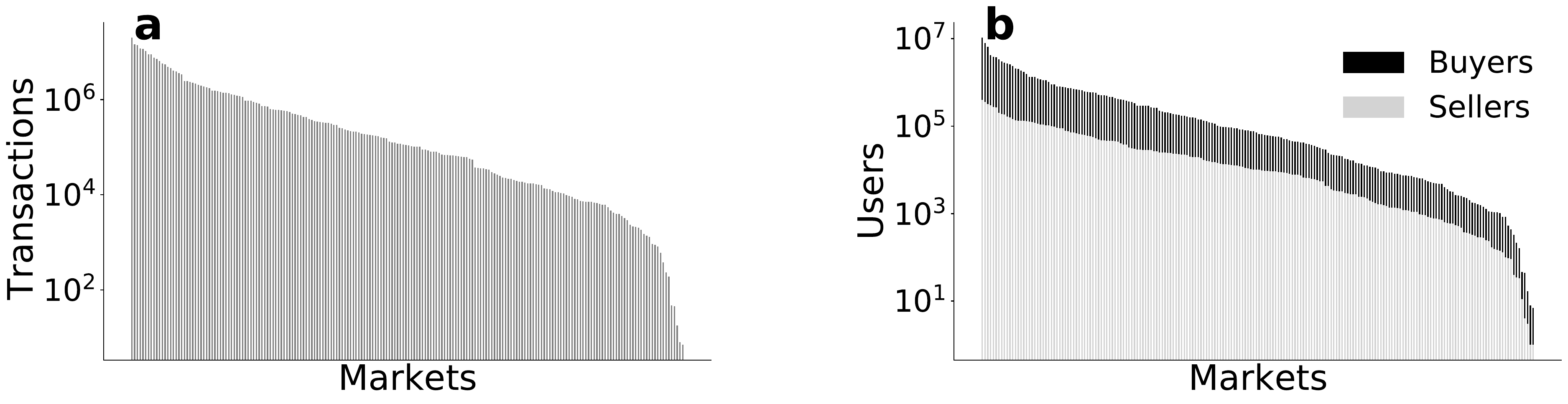}
	\caption{\label{fig:size_histogram} \textbf{E-Commerce Platform Markets} a): Plot of the number of transaction per each market of the e-commerce platform data. 
	b): number of buyers and sellers in each market of the e-commerce platform data.}
\end{figure*}

In Fig.~\ref{fig:size_histogram} we show the size of the 144 product markets in the regulated e-commerce platform dataset. 
In Fig.~\ref{fig:size_histogram}(a), we show the total number of transactions in each product market, whereas in Fig.~\ref{fig:size_histogram}(b) we show the total number of users (buyers and sellers). 
Product markets are heterogeneous in size, both w.r.t. number of transactions and number of users.

% \FloatBarrier

\pagebreak

% \FloatBarrier

% \FloatBarrier

% \FloatBarrier

% \FloatBarrier

% \FloatBarrier

% \FloatBarrier

% \FloatBarrier

% \FloatBarrier

% \FloatBarrier

\clearpage

% \FloatBarrier

\pagebreak 

\section*{Supplementary Tables}

In Tab.~\ref{tab:fit_coefficients} we show the fitted coefficients of $\beta$ and $c$ for every buyer degree class. 
The coefficients are slightly different among the different degree classes, but of the same order of magnitude, allowing us to fix the value of $beta$ and $c$ in the model simulation.
This is only an approximation in the context of a model whose goal is to see the role of different mechanisms in determining the structure and evolution of the buyer-seller network. 
If the goal was to reproduce the finest detail of the network, or to make predictions on its evolution, we'd assign values of $\beta$ and $c$ according to the buyer degree class, where the degree is sampled from the data degree distribution.

In Tab.~\ref{tab:dwm_numbers} we show details on each DWM in the dataset.
Details include the name,  start and end data, reason of closure, type of goods traded, total number of transactions and total volume of transactions in USD.

In Tab.~\ref{tab:delta} we show the value of the preferential attachment parameter $\mu$ fitted for each product market. 
While values are heterogeneous, showing the different role of preferential attachment in each market, the precise value is not important in our study.
Indeed, our only goal is to reproduce the main stylized facts of the data, not to reproduce the finest details of the network, and therefore changing the value of $\mu$ around the fitted value would not change our conclusions.

\begin{table}\centering
\caption{\label{tab:fit_coefficients} Memory kernel fitted coefficients $\beta$ and $c$ for every degree class  $k_{min} < $degree$< k_{min}*2$}
\begin{tabular}{rrr}
\toprule
 
$k_{min}$ &  $ \beta$ &     c \\
\midrule
         2 &    0.066 & 0.001 \\
         4 &    0.056 & 0.001 \\
         8 &    0.072 & 0.005 \\
        16 &    0.091 & 0.012 \\
        32 &    0.103 & 0.017 \\
        64 &    0.101 & 0.011 \\
       128 &    0.093 & 0.006 \\
\bottomrule
\end{tabular}
\end{table}
% \FloatBarrier

\begin{table*}\centering
\caption{\label{tab:dwm_numbers} Details of the DWMs under study: start, end, reason of closure, type, total number of transactions and total volume}
\begin{tabular}{lllllrr}
\toprule
                          Name &       Start &         End &    Closure &     Type &      \#Trx &  Volume [USD] \\
\midrule
         Silk Road Marketplace &  2011-01-31 &  2013-10-02 &     raided &    mixed &   840,987 &   131,604,274 \\
             Sheep Marketplace &  2013-02-28 &  2013-11-29 &       scam &    drugs &    65,904 &    10,923,327 \\
             Black Bank Market &  2013-10-18 &  2015-05-18 &       scam &    mixed &    89,444 &    13,152,830 \\
            Pandora OpenMarket &  2013-10-20 &  2014-11-05 &     raided &    drugs &    73,127 &     8,401,191 \\
                  Dream Market &  2013-11-01 &  2019-04-30 &  voluntary &    mixed &   570,734 &    57,637,323 \\
            Silk Road 2 Market &  2013-11-06 &  2014-11-05 &     raided &    mixed &   426,277 &    66,825,593 \\
                  Agora Market &  2013-12-03 &  2015-09-01 &  voluntary &    mixed &   911,094 &   141,473,388 \\
              Evolution Market &  2014-01-14 &  2015-03-19 &       scam &    drugs &   372,822 &    47,578,872 \\
      Middle Earth Marketplace &  2014-06-22 &  2015-11-04 &       scam &    mixed &    67,630 &     8,361,143 \\
 Russian Anonymous Marketplace &  2014-08-29 &  2017-07-15 &     raided &    mixed & 1,109,126 &    80,478,841 \\
               AlphaBay Market &  2014-09-01 &  2017-07-05 &     raided &    mixed & 4,263,740 &   546,010,808 \\
          Joker's Stash Market &  2014-10-07 &  2021-02-03 &    closed  &  credits &   998,687 &   153,138,403 \\
                Nucleus Market &  2014-10-24 &  2016-04-13 &       scam &    mixed &   391,394 &    56,594,214 \\
                Abraxas Market &  2014-12-24 &  2015-11-05 &       scam &    drugs &   168,642 &    21,854,042 \\
          Flugsvamp Market 2.0 &  2015-04-20 &  2018-10-02 &     closed &    drugs &   254,972 &    23,013,741 \\
                         Unicc &  2015-06-13 &  2021-02-10 &     active &  credits & 2,930,842 &   147,814,198 \\
                  Hansa Market &  2015-07-01 &  2017-07-20 &     raided &    drugs &   617,414 &    60,644,436 \\
             Hydra Marketplace &  2015-11-25 &  2021-02-10 &     active &    mixed & 6,005,608 & 2,175,558,739 \\
            Wall Street Market &  2016-09-09 &  2019-05-03 &     raided &    mixed &   681,825 &    48,153,667 \\
             TradeRoute Market &  2016-11-22 &  2017-10-12 &       scam &    mixed &   137,722 &    16,969,504 \\
                   Bypass Shop &  2017-03-10 &  2020-12-27 &     closed &  unknown & 1,041,438 &    65,663,561 \\
          Buybest - GoldenShop &  2017-11-13 &  2020-03-19 &     closed &  unknown &   386,046 &    24,449,110 \\
                 Silk Road 3.1 &  2018-02-10 &  2019-08-27 &       scam &    drugs &    93,426 &     9,053,684 \\
                 Empire Market &  2018-04-01 &  2020-08-30 &       scam &    mixed &   454,473 &   154,457,692 \\
                Apollon Market &  2018-05-03 &  2020-01-27 &       scam &    drugs &   106,395 &    12,902,953 \\
          Flugsvamp Market 3.0 &  2018-12-17 &  2021-02-10 &     active &  unknown &   291,018 &    39,344,294 \\
                        FEshop &  2019-08-14 &  2021-02-10 &     active &  unknown & 1,342,574 &    64,666,841 \\
                    DarkMarket &  2020-02-04 &  2021-01-12 &     raided &  unknown &   363,825 &    27,246,084 \\
\bottomrule
\end{tabular}
\end{table*}
% \FloatBarrier

\begin{table}\centering
\caption{\label{tab:delta}\textbf{Preferential attachment parameter $\Delta$.} Values of the preferential attachment parameter $\Delta$ for each product market, fitted with maximum likelihood estimation on the attractiveness distribution.}
\begin{tabular}{lr}
\toprule
Market & $\Delta$        \\
\midrule
1      &       85 \\
2      &       21 \\
3      &      155 \\
4      &      225 \\
5      &      400 \\
6      &      220 \\
7      &       90 \\
8      &       14 \\
9      &       50 \\
10     &       27 \\
11     &      290 \\
12     &       15 \\
13     &       13 \\
14     &       70 \\
15     &       33 \\
16     &       14 \\
17     &      175 \\
18     &      140 \\
19     &      240 \\
20     &      210 \\
21     &      180 \\
22     &      200 \\
23     &       65 \\
24     &       35 \\
25     &      190 \\
26     &      410 \\
27     &      220 \\
28     &      230 \\
\bottomrule
\end{tabular}
\end{table}

\end{document}